\documentclass[preprint,5p,twocolumn,numbers]{elsarticle}

\usepackage{amssymb}
\usepackage{amsmath}
\usepackage[colorlinks=true, linkcolor=blue, citecolor=blue, urlcolor=blue]{hyperref}
\usepackage{balance}
\usepackage{graphicx}
\usepackage{booktabs}
\usepackage{multirow}
\usepackage{tabularx}
\usepackage{enumitem}
\usepackage{threeparttable}
\usepackage{url}

\DeclareMathOperator{\E}{E}
\DeclareMathOperator{\Var}{Var}
\DeclareMathOperator{\RQ}{RQ}
\DeclareMathOperator{\Model}{M}
\DeclareMathOperator{\AdjModel}{\tilde{M}}

\DeclareMathOperator{\Degree}{Deg}
\DeclareMathOperator{\Continuous}{\mathcal{C}}
\DeclareMathOperator{\Dichotomous}{\mathcal{D}}

\newcommand{\vc}[1]{\boldsymbol{#1}}

\newcommand{\vczero}{\vc{0}}
\newcommand{\vcbeta}{\vc{\beta}}
\newcommand{\vctau}{\vc{\tau}}
\newcommand{\vcepsilon}{\vc{\epsilon}}
\newcommand{\mti}{\textup{\textbf{\textrm{I}}}}
\newcommand{\mtv}{\textup{\textbf{\textrm{V}}}}
\newcommand{\mtx}{\textup{\textbf{\textrm{X}}}}
\newcommand{\vcx}{\textup{\textbf{\textrm{x}}}}
\newcommand{\vcy}{\textup{\textbf{\textrm{y}}}}

\binoppenalty=\maxdimen
\relpenalty=\maxdimen

\newcommand{\change}[1]{\textcolor{black}{#1}}

\begin{document}

\title{A Case Study on Software Vulnerability Coordination}
\author[utu]{Jukka Ruohonen\corref{cor}}
\ead{juanruo@utu.fi}
\author[utu]{Sampsa Rauti}
\author[utu,pori]{Sami Hyrynsalmi}
\author[utu]{Ville Lepp\"anen}
\cortext[cor]{Corresponding author.}

\address[utu]{Department of Future Technologies, University of Turku, FI-20014 Turun yliopisto, Finland}
\address[pori]{Pori Department, Tampere University of Technology, P.O.\ Box 300,  FI-28101 Pori, Finland}

\begin{abstract}
\textit{Context}: Coordination is a fundamental tenet of software engineering. Coordination is required also for identifying discovered and disclosed software vulnerabilities with  Common Vulnerabilities and Exposures (CVEs). Motivated by recent practical challenges, this paper examines the coordination of CVEs for open source projects through a public mailing list.
\flushleft
\textit{Objective}:
\change{T}he paper observes the historical time delays between the assignment of CVEs on a mailing list and the later appearance of these in the National Vulnerability Database (NVD). Drawing from research on software engineering coordination, software vulnerabilities, and bug tracking, the delays are modeled through three dimensions: social networks and communication practices, tracking infrastructures, and the technical characteristics of the CVEs coordinated.
\flushleft
\textit{Method}: Given a period between 2008 and 2016, a sample of over five thousand CVEs is used to model the delays with nearly fifty explanatory metrics. Regression analysis is used for the modeling.
\flushleft
\textit{Results}: The results show that the CVE coordination delays are affected by different abstractions for noise and prerequisite constraints. These abstractions convey effects from the social network and infrastructure dimensions. Particularly strong effect sizes are observed for annual and monthly control metrics, a control metric for weekends, the degrees of the nodes in the CVE coordination networks, and the number of references given in NVD for the CVEs archived. Smaller but visible effects are present for metrics measuring the entropy of the emails exchanged, traces to bug tracking systems, and other related aspects. The empirical signals are weaker for the technical characteristics.
\flushleft
\textit{Conclusion}: Software vulnerability and CVE coordination exhibit all typical traits of software engineering coordination in general. The coordination perspective elaborated and the case studied open new avenues for further empirical inquiries as well as practical improvements for the contemporary CVE coordination.
\end{abstract}


\maketitle

\section{Introduction}

Software bugs have a life cycle.\footnote{~This paper is a rewritten and extended version of an earlier conference paper~\cite{Ruohonen17IWSMMensura} presented at IWSM Mensura 2017.} In a relatively typical life cycle, a bug is first introduced during development with a version control system, then reported in a bug tracking system, and then again fixed in the version control system. Also software security bugs, or vulnerabilities, follow a similar life cycle. Unlike conventional bugs, however, vulnerabilities often require coordination between multiple parties. Coordination is visible also during the identification and archiving of vulnerabilities with unique CVEs.

\begin{figure}[th!b]
\centering
\includegraphics[width=\linewidth, height=3.2cm]{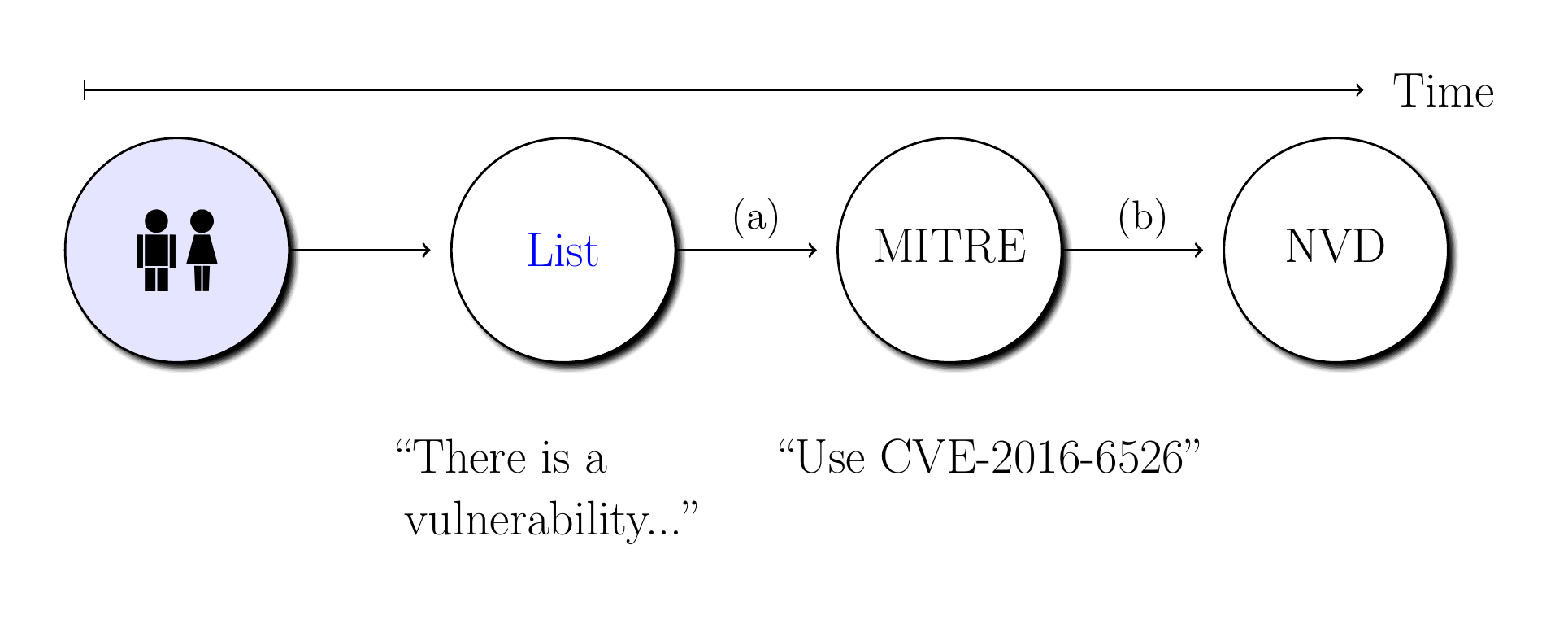}
\caption{Coordination via \texttt{oss-security} (2008 -- 2016)}
\label{fig: osec}
\end{figure}

There are four ways to obtain these universally recognized vulnerability identifiers. For obtaining a CVE, (a) an affiliation with an assignment authority (such as Mozilla or Microsoft) is required, but coordination may be done also by (b) contacting such an authority, making (c) a direct contact to the MITRE corporation, or (d) using alternative channels for public coordination \cite{Seifried17}. During the period observed, the public channel referred to the \texttt{oss-security} mailing list. The typical workflow on the list resembled the simple communication pattern illustrated in Fig.~\ref{fig: osec}.

Coordination is one of the major inhibitors of software development \cite{Johnson16a}. In the ideal case, the timeline from the leftmost circle would be short in terms of the timestamp that is recorded for a reported vulnerability to appear in NVD. In the future, robots might do the required coordination and evaluation work, but, recently, the time lags have allegedly been long. These delays have also fueled criticism about the whole tracking infrastructure maintained by MITRE and associated parties (for a recent media take see~\cite{Leyden17}). The alleged delays exemplify a basic characteristic of coordination: the presence of coordination often appears invisible to outsiders until coordination problems make it visible \cite{MaloneCrowston94}. The recent criticism intervenes with other transformations. These include the continuing prevalence of so-called vulnerability markets \cite{Ruohonen16RCIS}, including the increasingly popular crowd-sourcing bug bounty platforms~\cite{Laszka16, Ruohonen18WEIS}. Governmental interests would be another example \cite{Ablon17}. The CVEs assigned are commonly used also to track exploitable vulnerabilities on criminal underground platforms~\cite{Allodi17c}. A further example would be the use of social media and the resulting information leakages of the sensitive information coordinated~\cite{Leyden17, Sabottke15, Syed18}. These and other factors have increased the volume of discovered vulnerabilities for which CVEs are requested.

Reflecting these challenges, the coordination of CVEs through the mailing list ended in February 2017. To some extent, it is reasonable to conclude that \texttt{oss-security} lost its appeal as an efficient coordination and communication medium for CVE identifiers. To answer to the challenges, new options were provided for public CVE tracking \cite{Openwall17a}, and larger reforms were implemented at MITRE.

Motivated by these practical challenges, the paper examines the CVE coordination through the \texttt{oss-security}  mailing list between 2008 and 2016. The paper continues the earlier case study \cite{Ruohonen17IWSMMensura} by explicitly focusing on the timeline (b) in Fig.~\ref{fig: osec}. In other words, the present paper observes the time between CVE assignments on the mailing list and the later publication of these identifiers in NVD. Although the previous study revealed interesting aspects about social networks for open source CVE coordination, it also hinted that the coordination delays might be difficult to model and predict. Modeling of the delays is the goal of the present paper. To empirically model the delays, nearly fifty metrics are derived in this paper for proxying three dimensions of coordination: (i)~social networks and communication practices, (ii) tracking infrastructures within which the vulnerabilities had often already appeared prior to \texttt{oss-security}, and (iii) the technical characteristics of the vulnerabilities coordinated. Before elaborating these dimensions in detail, the opening Section~\ref{section: background} discusses the background and introduces the case studied. The research approach used to study the case is described in Section~\ref{section: approach}. Results are presented in Section~\ref{section: results} and further discussed in Section~\ref{section: discussion}.

\section{Background}\label{section: background}

Software engineering coordination is a fundamental part of the grand theories in the discipline \cite{Johnson16a}. The following discussion will briefly motivate the general background by considering some of the basic coordination themes in terms of software vulnerabilities and their coordination. After this motivation, the case studied is elaborated from a theoretical point of view. The discussion ends to a formulation of three research questions for the empirical analysis.

\subsection{Vulnerability Coordination}

Coordination can be defined as a phenomenon that originates from \textit{dependencies} between different \textit{activities}, and as a way to \textit{manage} these dependencies between the activities \cite{HowisonCrowston14}. Software engineering is team work. As team work implies social dependencies between engineers, there is always some amount of coordination in all software engineering projects involving two or more engineers. There are also technical dependencies in all software products. Early work on software engineering coordination focused on the reduction of such technical dependencies in order to improve task allocation and work parallelism~\cite{Blincoe15}. However, not all technical dependencies can be eliminated. For this reason, later work has often adopted a \textit{socio-technical} perspective for studying software engineering coordination.

The augmentation of technical characteristics with social tenets is visible at a number of different fronts. One relates to the research questions examined. Examples include synchronization of work and milestones, incremental integration of activities, arrangement of globally distributed teams, frequent deliveries, and related project management processes \cite{Lee13, McChesney97, Paasivaara03}. Another relates to the research methodologies used. While interviews and surveys are still often used \cite{Blincoe15, Lee13, Paasivaara03}, techniques such as social network analysis have become increasingly common for examining software engineering coordination \cite{Bird11, Sierra18}. Despite of these changes, one fundamental theoretical premise has remained more or less constant over the years.

Coordination requires \textit{communication}, and coordination failures are often due to communication problems~\cite{Blincoe15, Wolf09b}. The consequences from coordination failures vary. Typical examples include schedule slips, duplication of work, build failures, and software bugs \cite{Bird11, Blincoe15}. Communication obstacles, coordination failures, and the resulting problems are well-known also in the security domain.

A good example would be incident and abuse notifications for vulnerable Internet domains: it is notoriously difficult to communicate the security issues discovered to the owners or maintainers of such vulnerable domains~\cite{CetinVanEeeten17}. Analogous problems have been prevalent in \textit{vulnerability disclosure}, which refers to the practices and processes via which the vulnerability discoverers make their discoveries known to the vendors whose products are affected either directly or via a third-party coordinator. Although there have long been  recommendations and guidelines~\cite{Christey02}, it is still today often difficult to communicate vulnerability discoveries to vendors \cite{Ring15}. A~substantial amount of effort may be devoted to coordinate the disclosure and remediation of high-profile vulnerabilities, but difficulties often occur for more mundane but no less important vulnerabilities. Some vendors are reluctant to participate in vulnerability disclosure---and some vendors even avoid patching their software products altogether. Even legal threats are still today not unheard of. While vulnerability disclosure is a prime example of coordination (failures) in the software security context, it has only a narrow scope.

As Steven M. Christey---the primary architect behind the current CVE tracking---has argued, the term \textit{vulnerability coordination} is preferable because vulnerability disclosure only captures a small portion of the activities required to handle and archive vulnerabilities \cite{Christey13}. In fact, vulnerability coordination is a relatively frequent activity among the integrative software engineering activities carried out by open source developers \cite{Adams16}. Bug reports must be handled in a timely manner for security issues. Evaluation is needed to assess the versioned products affected. Of course, also fixes must be written, but fixes may further require careful backtracking within version control systems, debugging, reviews from peers, testing done by peers and users alike, and coordination between business partners or third-party open source software projects. Then, erratas, security advisories, or other notes must be written, identifiers must be allocated to uniquely track the notes and to map these to other identifiers, preparations may be required for answering to user feedback, and so forth. These are all examples of the software engineering activities that are typically present in the software vulnerability context.

For empirical software engineering research, the identifiers used to track vulnerabilities are particularly relevant. The most important identifier is CVE, in terms of both research and practice. For open source projects, having a single canonical identifier helps developers and users to coordinate their efforts. Given the inconsistency issues affecting open bug trackers \cite{Romo14, TianHassan16}, CVEs help at addressing questions such as: ``well it sounds like this one but maybe it's that other one?'' \cite{Seifried17}. For a security professional, CVEs are ubiquitous entries in a curriculum vitae. For research, CVEs provide the starting point for connecting the distinct engineering activities into a coherent schema. While there is an abundant amount of existing research operating with CVE-based research schemas, thus far, only one attempt \cite{Ruohonen17IWSMMensura} has been made to better understand the primary schema; software and security engineering coordination is required for CVEs themselves.

\subsection{CVE Coordination Through a Mailing List}

Software development teams tend to produce software products that reflect the communication structure of the teams, to paraphrase the famous law formulated by Melvin E. Conway~\cite{Conway68}. Although the law is not directly applicable to the context of CVE coordination, the \texttt{oss-security} case supports a weaker variant of the same basic assertion. This variant might be defined as a statement that a communication structure tends to reflect the \textit{type} of software engineering work and the \textit{medium} used for communication.

The type of work done and the  medium both characterize also the social network structure of the CVE coordination through the mailing list. Both limit also the applicability of common theoretical interpretations. For instance, knowledge sharing is a typical way to  motivate research on open source social networks \cite{Kuk06, Licorish14, Toral09}, but sharing of \textit{knowledge} was not the main purpose of the \texttt{oss-security} mailing list during the period studied.

The primary purpose of the list was to coordinate the assignment of CVE identifiers for discovered and usually already disclosed software vulnerabilities. This coordination was done by explicitly requesting CVE identifiers, which were then assigned by MITRE affiliates based on a brief evaluation. In terms of communication practices, this coordination practice often culminated in short replies, such as ``use CVE-2016-6526'' \cite{Openwall16c}, made by MITRE affiliates to previously posted CVE requests. Participants typically kept MITRE's \texttt{cve-assign@mitre.org} email address in the carbon copy field when they posted a request, further using a subject line that identified the message as a request. While longer discussions were not unheard of, these communication practices indicate that knowledge sharing has only limited appeal for framing the case theoretically. Exchanging information about abstract identifiers does not necessarily imply thorough discussions about the technical details of the vulnerabilities coordinated. In other words, data does not equate to information, and information does not equal knowledge.

\begin{figure}[th!b]
\centering
\includegraphics[width=\linewidth, height=6cm]{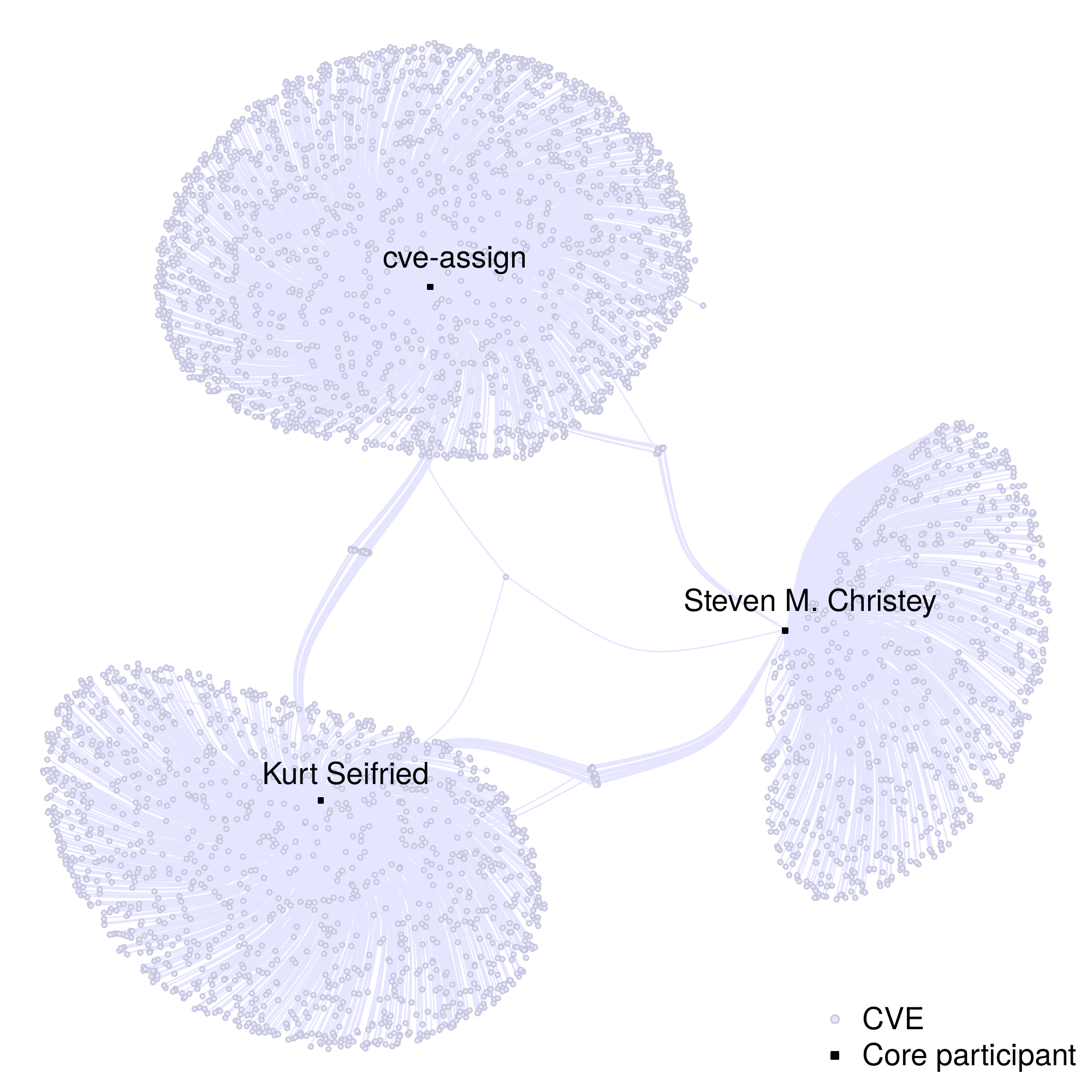}
\caption{The Core Participants (one hop from the labeled vertices)}
\label{fig: core}
\end{figure}

The coordination work through the list produced clear core social network components \cite{Ruohonen17IWSMMensura}. This observation is classical in the open source context \cite{Conaldi12, Crowston06, Licorish14, Toral10}, but the contextual interpretation is still relevant. Due to the way CVEs are assigned, the cores centered to MITRE affiliates. By using a network representation described later in detail, the core social network components are illustrated in Fig.~\ref{fig: core}. Excluding the use of MITRE's email address without an explicit sender, there are two identifiable individual participants with degrees higher than $800$, meaning that these two participants both coordinated at least eight hundred CVEs each. Both participants are MITRE affiliates. What is more, there are only a few CVEs that link the three core participants to each other. This observation indicates that task allocation and related coordination techniques apply also to the case studied.

These core social network components are a good example of so-called communication brokers who help at resolving coordination obstacles in software engineering~\cite{Wolf09a}. Such brokers integrate information from multiple sources. This \textit{integrative} engineering work often leads to a social network structure with relatively high level of centrality, low level of clustering, and star-like centers within which the integrative work is located~\cite{Parraguez15}. This theoretical characterization applies well also to the social networks for the CVE coordination through the mailing list.

\begin{figure*}[t!]
\centering
\includegraphics[width=14cm, height=6.5cm]{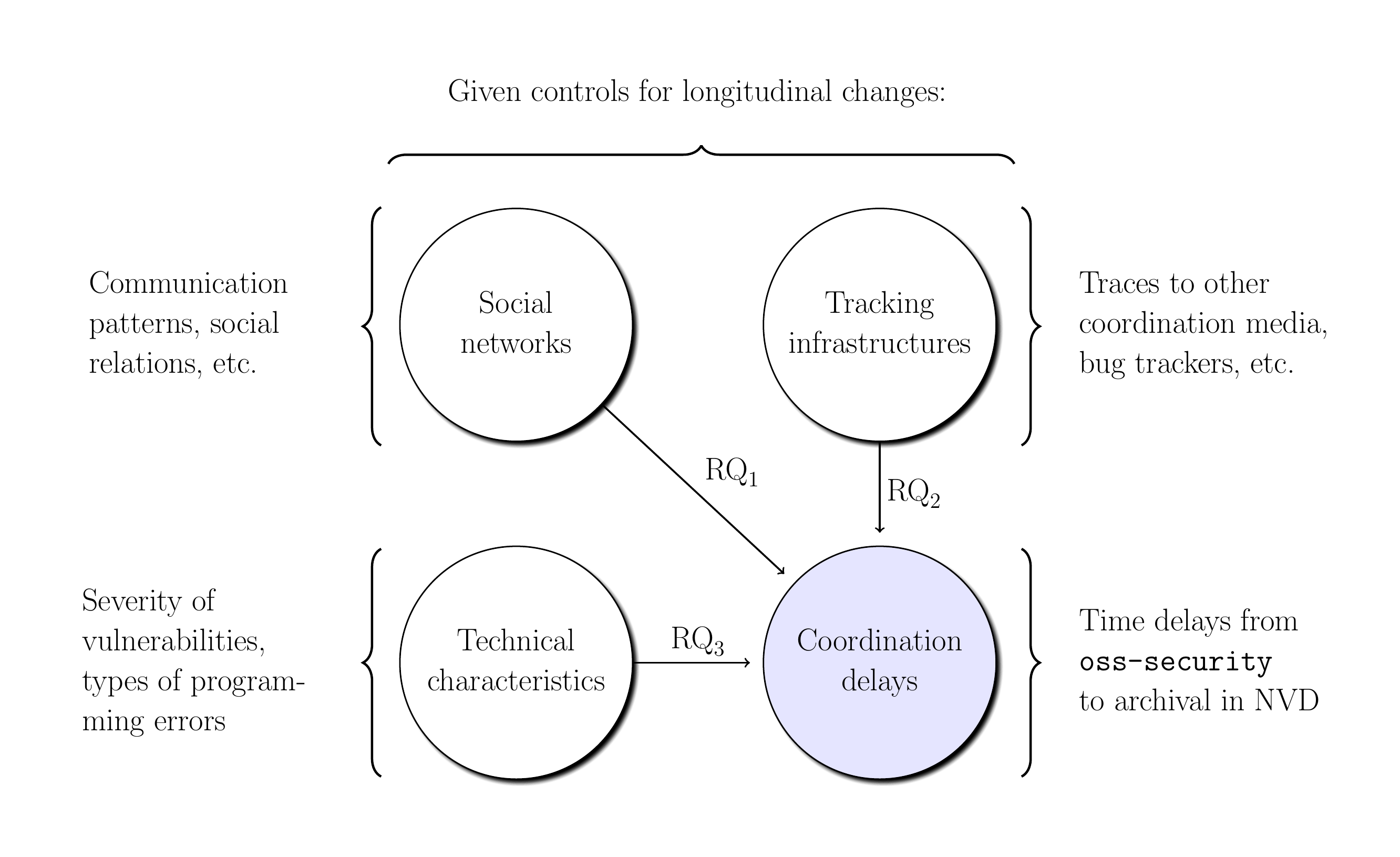}
\caption{Hypothetical Relations and the Corresponding Research Questions}
\label{fig: hypotheses}
\end{figure*}

Thus, the type of software engineering work carried out is an important factor characterizing the corresponding social network structures. In terms of development mailing lists, non-core participants may post as many messages as core participants~\cite{Crowston17}, but this observation does not generalize to all cases \cite{Guzzi13}, including the case studied. Although coordination requires communication, coordination tends to result in different network structures than communication required for other work (or leisure) activities. A~communication medium used for coordination presumably also affects the emerging network structure. For instance, social networks constructed from bug tracking systems indicate that many individuals may report bugs even though the actual development may be concentrated to a small core group~\cite{Zanetti13a}. A communication medium also places restrictions over who can participate, how much can be communicated, and what is communicated~\cite{Bird11, PooCaamano17}. While \texttt{oss-security} is open for anyone to participate, the communication volume and the content of messages exchanged are both important for further elaborating the case.

The integrative coordination work done on the list reflected not only social but also loose technical dependencies between different tracking infrastructures. The classical consumer-producer abstraction for software engineering coordination \cite{MaloneCrowston94, McChesney97} is useful for framing these technical dependencies theoretically. According to this abstraction, there exists a \textit{prerequisite constraint}: the work activity of a producer must be usually completed before work can start on the consumer side~\cite{HowisonCrowston14, MaloneCrowston94}. In terms of the case studied, the participants (producers) who requested CVEs provided sufficient technical information to justify the requests for the MITRE affiliates (consumers). If the information was sufficient for a request, the prerequisite constraint was satisfied, and the vulnerability in question later appeared in NVD with the CVE requested. Instead of in-depth knowledge sharing, the sufficient information was usually delivered via hyperlinks to other open source tracking infrastructures within which a given vulnerability had already been discussed. Consequently, the discovery, disclosure, and patching of the vulnerabilities coordinated had almost always occurred before information appeared on the mailing list.

\subsection{Research Questions}

The preceding discussion motivates three questions worth asking about the delays between CVE requests on the mailing list and the later appearance of these identifiers in the central tracking database. Given the medium used and the type of software engineering coordination done, these questions can be framed by separating the two terms in the concept of socio-technical coordination. The first question addresses the social dimension:

\begin{enumerate}[label={RQ$_{\arabic{enumi}}$}, resume]
\item{\textit{Have the CVE coordination delays been affected by the social networks and communication practices between the participants on the \texttt{oss-security} mailing list?}}\label{rq: social}
\end{enumerate}

The remaining two questions address the technical dimension in socio-technical coordination. Given the integrative work done and the consumer-producer abstraction, it is worth asking the following question about the delays:

\begin{enumerate}[label={RQ$_{\arabic{enumi}}$}, resume]
\item{\textit{Did traces to other tracking infrastructures affect the coordination delays during the period observed?}}\label{rq: infrastructure}
\end{enumerate}

The third and final research question approaches the technical dimension from a more direct perspective:

\begin{enumerate}[label={RQ$_{\arabic{enumi}}$}, resume]
\item{\textit{Were the coordination delays affected by the technical characteristics of the vulnerabilities coordinated?}}\label{rq: technical}
\end{enumerate}

The analytical meaning behind the three questions is illustrated in Fig.~\ref{fig: hypotheses}. It should be noted that causal inference is not attempted; hence, no arrows are drawn in the figure between the three explanatory dimensions.

\section{Approach}\label{section: approach}

In what follows, the research approach taken to study the CVE coordination delays is elaborated by introducing the empirical dataset, the operationalization of the delays, and the construction of the social networks observed. After this machinery has been installed, the approach is further elaborated by describing the explanatory metrics and the statistical methodology used to model the delays.

\subsection{Data}\label{subsec: data}

The dataset is compiled from two sources. All email messages posted on \texttt{oss-security} were obtained from the online Openwall archive \cite{Openwall16a}. These messages are cross-referenced with CVE identifiers to the second data source, NVD~\cite{NVD17a}. The sampling period runs from the first welcome message in February 2008 to the emails posted in December 31 2016. This sampling interval corresponds with the historical period during which the mailing list was used for CVE assignments.

\begin{figure}[th!b]
\centering
\includegraphics[width=\linewidth, height=5cm]{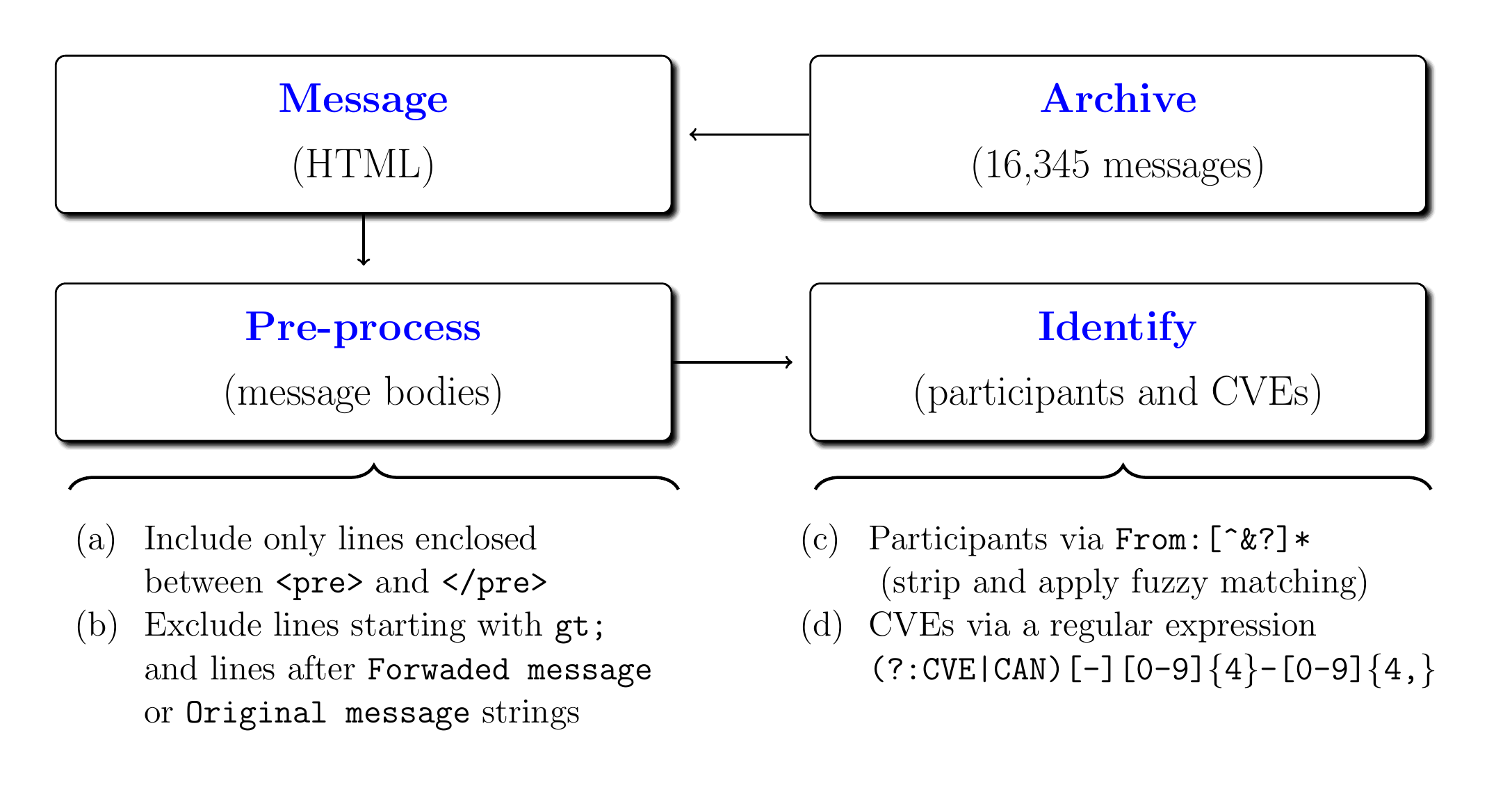}
\caption{Pre-processing in a Nutshell}
\label{fig: pre-processing}
\end{figure}

There were a little over sixteen thousand messages posted during this time interval. For each message delivered in the hypertext markup language (HTML) format, a twofold routine is used for pre-processing the archival data (see Fig.~\ref{fig: pre-processing}). The first pre-processing subroutine extracts the actual messages from the HTML markup, further omitting all lines referring to quotations from previous messages posted on the mailing list. To some extent, also forwarded messages are excluded based on simple but previously used heuristics~\cite{Wu05}. Due to the markup language, this pre-processing routine is likely to yield some inconsistencies. However, the consequences for the empirical analysis should be relatively small because no attempts are made to pre-process email threads. This choice is largely imposed by the use of the online archive. In particular, the fields \texttt{Message-ID}, \texttt{In-Reply-To}, and \texttt{References} are not delivered via the online interface, which prevents the use of common (meta-data) techniques for parsing email threads~\cite{Guzzi13, Wang14}. Although content-based alternatives exist for reconstructing threaded email structures \cite{Dehghani12, Schmid14}, only the senders of emails are considered in order to improve data quality and to simplify the data processing.

The second subroutine is used for identification. The \texttt{From} email header field is used for identifying the individual participants by using the first match from a simple Python regular expression. Although the matching itself is simple, it should be emphasized that all senders are identified according to their names rather than their addresses. This option is often preferred because individuals' email addresses tend to vary \cite{Bird11, Nia10, Toral09}. The choice is again also imposed by the online archive, which obfuscates the email addresses for spam prevention and related reasons. Consequently, additional pre-processing techniques \cite{Ngamkajornwiwat08} for mapping identified names to addresses cannot be used.

Furthermore, Levenshtein's \cite{Levenshtein66} classical distance metric between two names, say $L(x, y)$, is a decent choice for accounting small inconsistencies \cite{Zangerle13}. If $s_1$ and $s_2$ denote two full names with lengths $l_1$ and $l_2$, a similarity score is computed through $\delta = 1.0 - L(s_1, s_2) / \max\lbrace l_1, l_2 \rbrace$~\cite{Christen12}. If the scalar $\delta$ then exceeds a threshold value $0.8$, two names are taken to refer to the same individual. Although the threshold is subjective, it captures some typical cases, such as ``John Doe'' who occasionally writes his name as ``John, Doe'' when sending emails. The few outlying cases that exceeded the $\delta = 0.8$ threshold were further evaluated manually. This additional check revealed no obvious approximation mistakes. In addition, ``Christey, Steven M.'' and ``Steven M. Christey'' were merged manually.

Also CVEs are identified with a regular expression, which is a typical approach for searching vulnerabilities from heterogeneous sources~\cite{Allodi17c, LinaresVasquez17}. The expression takes into account both the old and deprecated candidate (CAN) syntax as well as the recently made syntax change that allows an arbitrary amount of digits in the second part of CVE identifiers \cite{MITRE15a, MITRE15b}. Even though forwarded messages and direct quotations to previous emails are approximately excluded, it should be noted that there can be many-to-many relations between messages and CVEs. For instance, a lengthy security advisory posted on the list may contain a large amount of individual CVE-referenced vulnerabilities. Such cases are included in the sample. Finally, only those CVEs are qualified that have also valid entries in NVD. Most of the invalid entries in the database refer to identifiers that were assigned but which were later rejected from inclusion to the database. Given that these disqualified CVE assignments cause different biases \cite{Christey13}, all rejected identifiers were excluded from the empirical sample. The matching of these invalid entries was done by searching for the string \texttt{REJECT} in the summary field provided by NVD.

\subsection{Coordination Delays}

There are multiple ways to measure the efficiency of software engineering coordination~\cite{Lee13}. In the context of software bugs in general, a good example would be the validity of bug reports. Bug tracking infrastructures typically reserve multiple categories for classifying invalid reports. These categories include classes for already fixed bugs, irreproducible bugs, and duplicate reports, among other groups. If a large amount of bug reports end up into such classes, coordination would be generally inefficient.

As coordination deals with dependencies, it is no surprise that also social network structures of bug reporters have been observed to affect the probability of valid bug reports~~\cite{Zanetti13a}. Although the generalizability of this observation seems limited \cite{Bhattacharya11}, there are good reasons to suspect that an analogous effect is present in the software vulnerability context. Because attribution is particularly important for vulnerabilities and monetary compensations are relatively common, it is no wonder that invalid---or even outright fake---reports have been a typical menace affecting vulnerability coordination~\cite{Christey13, Laszka16}. For this reason, vulnerability reports from established discoverers likely increase the validity of the reports as well as the trust placed on the reports. This trust provision is also reflected on the CVE assignment authorities granted for a few individual security researchers.

Due to the sensitiveness of the information coordinated, the availability of open data is limited about the internal vulnerability tracking systems used by software vendors, MITRE, and other actors. Many open source projects also limit the visibility of security bugs, or otherwise try to constrain the exposure of public information.

These data limitations have affected also the ways to quantify longitudinal vulnerability information. For instance, the efficiency of vulnerability disclosure can be approximately measured with a time difference between a disclosure notification sent to a software vendor and the vendor's reply \cite{Arora10, Ruohonen16AICCSA}. Analogously, patch release delays can be measured by fixing the other endpoint to the dates on which vendors released patches for the vulnerabilities disclosed to the vendors~\cite{McQueen09, Temizkan12}. Similar delays can be measured also in terms of initial bug reports and later CVE assignments based on the reports~\cite{Wijayasekara12}. Further examples include the timing of security advisories~ \cite{Ruohonen17COMSIS}, the dates on which signatures were added to intrusion detection and related systems~\cite{Bilge12}, and exploit release dates for known vulnerabilities~\cite{Ablon17, Allodi17c, Bozorgi10}. All these different dates convey different viewpoints on the coordination of vulnerabilities.

In this paper, analogously, the empirical interest relates to the following per-CVE time differences (in days):
\begin{equation}\label{eq: delays}
y_i =
T_{\textmd{NVD}_i} -
T_{\textmd{\texttt{oss-security}}_i} ,
\end{equation}
given
\begin{equation}\label{eq: delays restriction}
T_{\textmd{NVD}_i} \geq
T_{\textmd{\texttt{oss-security}}_i}
\quad\textmd{for all}\quad i = 1, \ldots, n
\end{equation}
vulnerabilities observed. The timestamp $T_{\textmd{NVD}_i}$ records the date on which the $i$:th CVE was first stored to NVD. The second timestamp $T_{\textmd{\texttt{oss-security}}_i}$ refers to the \textit{earliest} date on which this CVE was posted on the mailing list. The restriction in \eqref{eq: delays restriction} excludes already archived CVEs that were later discussed on the mailing list. Thus, the integer $y_i \geq 0$ approximates the length of the path (b) in Fig.~\ref{fig: osec}. In theory, also the path (a) in Fig.~\ref{fig: osec} could be measured, but the data from the online archive makes it difficult to identify the initial CVE requests. This limitation does not affect the validity of the delay metric, but it does affect the interpretation given for it.

The metric in \eqref{eq: delays} approximates the time delays between the initial CVE assignments done through the \texttt{oss-security} mailing list and the usually later appearance of the requested CVEs in NVD. Therefore, $y_i$ focuses on the \textit{internal} coordination done by MITRE affiliates, the NVD team, and other actors involved in the coordination. One way to think about this internal coordination is to consider the delay metric as an \textit{indirect} quantity for measuring how fast work items in a \textit{backlog} or an \textit{inventory} are transferred to complete deliveries \cite{MaloneCrowston94, Ruohonen19ACI}. Due to data limitations, however, it is currently neither possible to observe the internal within-MITRE (or within-NVD) coordination directly nor to explicitly measure the work items in the CVE backlog. Therefore, the delay metric used provides a sensible but not entirely reliable way to approximate the typical coordination delays that affected one particular CVE coordination channel.

\subsection{Bipartite Email and Infrastructure Networks}

\begin{figure}[th!b]
\centering
\includegraphics[width=\linewidth, height=4.5cm]{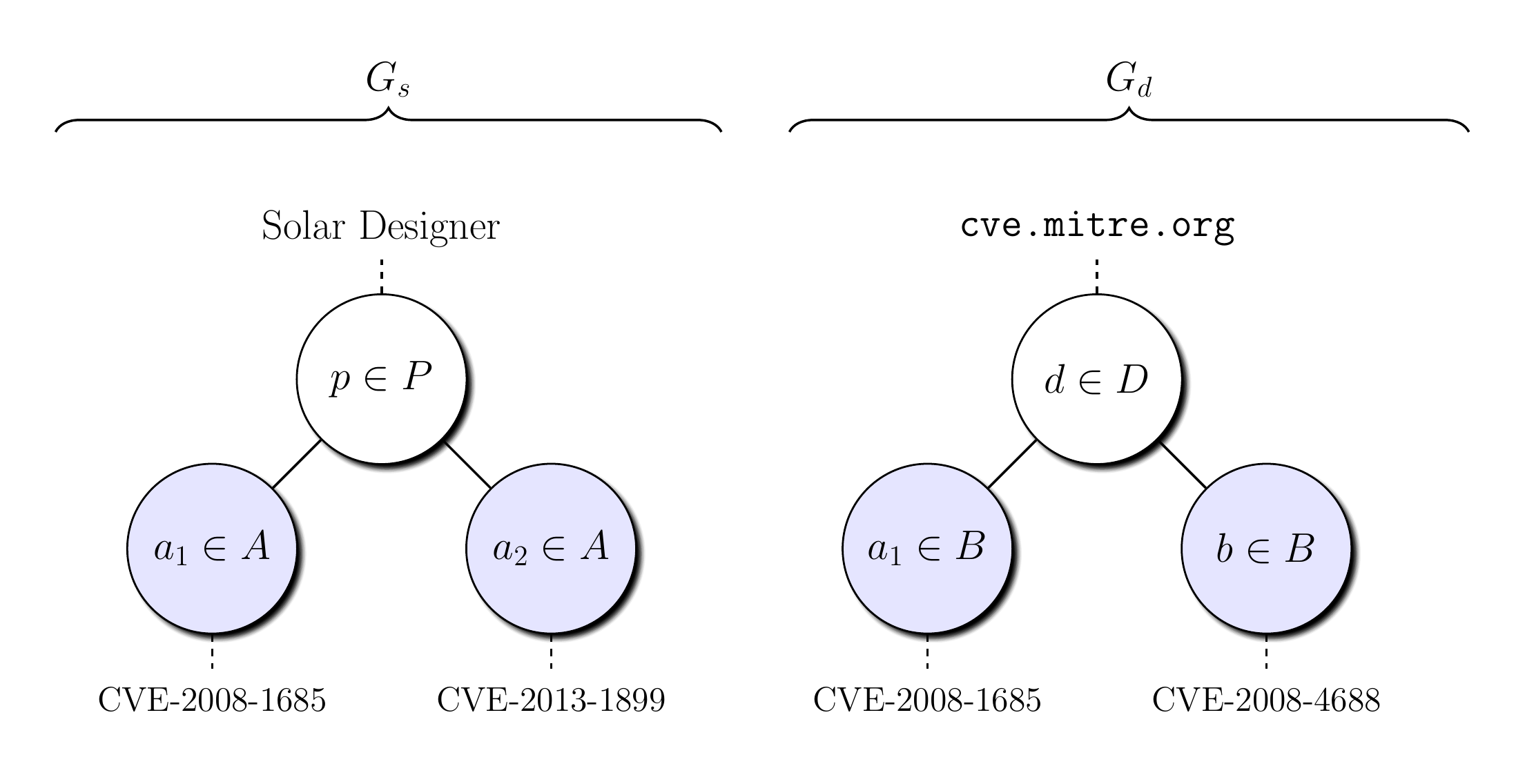}
\caption{Example Networks}
\label{fig: example networks}
\end{figure}

The socio-technical coordination and communication characteristics are proxied by observing so-called task-based~\cite{Wolf09a} or person-task \cite{Sierra18} social networks. Thus, the underlying social network structure is bipartite, meaning that there are two types of vertices. An edge connecting any two vertices always contains both types; there are no edges that would connect a vertex of one type to a vertex of the same type. In addition, the network structure is unweighted and undirected. More formally:
\begin{equation}\label{eq: social network}
G_s = (P \cup A, E) ,
\end{equation}
where $G_s$ denotes an observed social network constructed from the messages that were posted on the mailing list, from the first message posted in 2008 to the last message in 2016. The two disjoint vertex sets $P$ and $A$ refer to participants and CVE identifiers, respectively. If a participant $p \in P$ has sent a message containing a CVE identifier $a \in A$, an undirected edge, $(p, a) = (a, p)$, is present in the edge set $E$. Therefore, individual participants are linked together through CVE identifiers (see Fig.~\ref{fig: example networks}). Due to the bipartite structure, it holds that $P \cap A = \emptyset$ and
\begin{equation}\label{eq: degrees}
\vert E \vert =
\sum_{p \in P}\Degree(p) =
\sum_{a \in A}\Degree(a) ,
\end{equation}
where
\begin{equation}\label{eq: deg}
\Degree(x) = \sum_{(x, y) \in E} y .
\end{equation}

Furthermore, another network is constructed from the uniform resource locators (URLs) embedded in the hyperlinks present in the emails that contained also CVEs. The structure is again undirected, unweighted, and bipartite:
\begin{equation}\label{eq: domain network}
G_d = (D \cup B, F) ,
\end{equation}
where $D$ denotes a set of domain names extracted from the URLs sent by participants in $P$, $B$ denotes another set of CVEs, and $F$ is an edge set containing edges from the domain names to the CVEs. If a participant $p \in P$ sent an email containing a $b \in B$ and a domain name $d \in D$ extracted from a URL in a hyperlink, an undirected edge $(b, d) = (d, b)$ is present in the set $F$. Thus, CVEs are linked also to domain names through the participants who posted messages containing both CVE identifiers and hyperlinks. Therefore, $B \subseteq A$ and $n = \vert A \vert \geq \vert B \vert$ because the subset of CVE identifiers stored to $B$ are required to also have mappings to domain names.

The network $G_s$ is a \textit{social} network in the traditional sense; even though participants are linked to each other through abstract identifiers, the participants are still \textit{human} beings. The network $G_d$, in contrast, resembles more the so-called domain name system graphs within which domain names are connected to each other via Internet protocol (IP) addresses or by other technical relations~\cite{Ruohonen17CIT, Stevanovic16}. Consequently, it would be possible to manipulate $G_d$ by resolving the addresses of the domain names or by considering only the second-level domain names \cite{Ruohonen16MESSA}. Given the historical context, however, no attempts are made to resolve the domain names, many of which are nonexistent today. Instead, only three semantic validation checks are enforced: (a) the length of each $d \in D$ is asserted to be at least three characters; (b) each $d$ is required to contain a dot character; (c) and no entry in $D$ is allowed to refer to a semantically valid IPv4 address.

\subsection{Explanatory Metrics}\label{subsec: explanatory metrics}

The three research questions are evaluated by regressing the coordination delays $y_1, \ldots, y_n$ against the metrics enumerated in Table~\ref{tab: metrics}. Six \textit{models} ($\Model$) are used for the statistical computations; the integer $k$ denotes the cumulative number of metrics included in the consecutively estimated models, including the intercept. The table shows also the scale of the metrics; there are a few \textit{continuous} ($\Continuous$) metrics but most of the metrics are \textit{dichotomous} ($\Dichotomous$) dummy variables. The metrics enumerated can be further elaborated according to the models within which these first appear.

\begin{table*}[t!]
\centering
\caption{Explanatory Metrics}
\label{tab: metrics}
\begin{threeparttable}
\begin{small}
\renewcommand*{\arraystretch}{1.2}
\begin{tabular}{lclcl}
\toprule
$\RQ_i$ & $\Model_j$ & Name & Scale & Description \\
\hline
-- & $\Model_1$ & $2009, \ldots, 2016$ & $\Dichotomous$ & True for CVEs assigned on a given year according to $T_\textmd{\texttt{oss-security}}$. \\
& $(k = 21)$ & Feb, $\ldots$, Dec & $\Dichotomous$ & True for CVEs assigned on a given month according to $T_\textmd{\texttt{oss-security}}$. \\
&& WEEKEND & $\Dichotomous$ & True for CVEs assigned on Saturday or Sunday according to $T_\textmd{\texttt{oss-security}}$. \\
\hline
$\RQ_1$ & $\Model_2$
& SOCDEG & $\Continuous$ & The degree of all CVE-labeled vertices in $G_s$. \\
& $(k = 25)$ & MITREDEV & $\Dichotomous$ & True for CVEs in the neighborhood of the three labeled vertices in Fig.~\ref{fig: core}. \\
&& MSGSLEN & $\Continuous$ & The amount of characters divided by $100$ in the emails mentioning a CVE. \\
&& MSGSENT & $\Continuous$ & For a given CVE, the Shannon entropy of the emails mentioning the CVE. \\
\hline
$\RQ_2$ & $\Model_3$
& INFDEG & $\Continuous$ & The degree of CVE-labeled vertices in $G_d$ (zero for any $b \in B$ but $b \not\in A$). \\
& $(k = 31)$ & NVDREFS & $\Continuous$ & Number of reference URLs given in NVD for the CVEs observed. \\
&& VULNINF & $\Dichotomous$ & True for CVEs linked to vulnerability infrastructures via $G_d$. \\
&& BUGS & $\Dichotomous$ & True for CVEs linked to bug tracking and related systems via $G_d$. \\
&& REPOS & $\Dichotomous$ & True for CVEs linked to version control and related systems via $G_d$.\\
&& SUPPORT & $\Dichotomous$ & True for CVEs linked to vendors' support channels via $G_d$. \\
\hline
$\RQ_3$ & $\Model_4$
& IMPC & $\Dichotomous$ & True for CVEs having a partial or a complete impact on confidentiality \\
& $(k = 34)$ & IMPI & $\Dichotomous$ & True for CVEs having a partial or a complete impact on integrity \\
&& IMPA & $\Dichotomous$ & True for CVEs having a partial or a complete impact on availability \\
\cmidrule{2-5}
& $\Model_5$ & EXPNET & $\Dichotomous$ & True for CVEs that may be exploited only with a network access. \\
& $(k = 37)$ & EXPCPLX & $\Dichotomous$ & True for CVEs with a high or a medium access complexity for exploitation. \\
& & EXPAUTH & $\Dichotomous$ & True for CVEs that can be exploited only through authentication. \\
\cmidrule{2-5}
& $\Model_6$ & CWE-264 & $\Dichotomous$ & True for CVEs in the domain of permissions, privileges, and access controls.\\
& $(k = 47)$ & CWE-119 & $\Dichotomous$ & True for CVEs in the domain of buffer-related bugs.\\
&& CWE-79 & $\Dichotomous$ & True for CVEs in the domain of cross-site scripting (XSS).\\
&& CWE-20 & $\Dichotomous$ & True for CVEs in the domain of input validation. \\
&& CWE-200 & $\Dichotomous$ & True for CVEs in the domain of information leaks. \\
&& CWE-399 & $\Dichotomous$ & True for CVEs in the domain of resource management bugs. \\
&& CWE-189 & $\Dichotomous$ & True for CVEs in the domain of numeric bugs. \\
&& CWE-352 & $\Dichotomous$ & True for CVEs in the domain of cross-site request forgery (CSRF).\\
&& CWE-89 & $\Dichotomous$ & True for CVEs in the domain of structured query language (SQL) injection. \\
&& CWE-310 & $\Dichotomous$ & True for CVEs in the domain of cryptographic bugs. \\
\bottomrule
\end{tabular}
\end{small}
\end{threeparttable}
\end{table*}

\subsubsection{Control Metrics}

Temporal aggregation of social network data should be done only after a careful consideration \cite{Nia10}. Previous work in the \texttt{oss-security} context also indicates that the social networks for the open source CVE coordination changed over the years \cite{Ruohonen17IWSMMensura}. In contrast to what has been claimed to characterize open source projects~\cite{Ngamkajornwiwat08}, the coordination effort did not diminish over time. In fact, the list became more popular, which resulted in more participants and more coordinated CVEs. These transformations also changed the social network structure, although the core of the network structure remained centered to MITRE affiliates. A notable change occurred also in the structure of this network core: Kurt Seifried from Red Hat joined the CVE editorial board~\cite{MITRE15c} and took an active role also on \texttt{oss-security}. This activity reduced the reliance on a single MITRE affiliate for the CVE coordination through the list, resulting in the network cores illustrated in Fig.~\ref{fig: core}.

Instead of explicitly modeling these changes through separate annual social networks---as has been typical in applied social network research \cite{Ruohonen16AICCSA, Ruohonen17IWSMMensura, Toral10, Zanetti13b}, the longitudinal dimension is approximately controlled with eight dummy variables. Each of these is zero for the $i$:th identifier unless the CVE identifier was assigned on a given year between 2009 and 2016 according to the corresponding $T_{\textmd{\texttt{oss-security}}_i}$ timestamp used in \eqref{eq: delays}. The initial year 2008 acts as the reference category against which the effects of these annual dummy variables are compared against.

Given the fairly complex time series dynamics arising from the archiving of vulnerabilities to NVD and related tracking infrastructures~\cite{JohMalaiya17, Tang19}, a further set of eleven dummy variables is included for controlling potential monthly variation in the coordination delays. The $T_{\textmd{\texttt{oss-security}}_1}, \ldots,
T_{\textmd{\texttt{oss-security}}_n}$ timestamps are again used for computation, and January acts as the reference
month.

The third and final longitudinal control metric is named WEEKEND. It scores a value one for a CVE posted to the mailing list on Saturday or Sunday according to the coordinated universal time, taking a value zero otherwise. The rationale relates to observations that the days of week may affect the likelihood of introducing bugs during software development~\cite{Sliwerski05}. Although the empirical evidence is mixed regarding this assertion \cite{Eyolfson11}, it is reasonable to extend it toward vulnerability coordination. For instance: if the $i$:th requested CVE would have otherwise ended up to NVD rapidly after two days, it may be that an additional delay, say $\epsilon > 0$, is present in case the request was posted on a weekend, such that $y_i = 2 + \epsilon$. The same rationale applies to the monthly effects. In other words, annual holidays presumably taken by the MITRE affiliates and others participants may well affect the coordination delays.

\subsubsection{Social Network and Communication Metrics}\label{subsec: social network and communication metrics}

Four metrics are used for soliciting an answer to \ref{rq: social}. The first is the amount of participants linked to CVEs:
\begin{equation}\label{eq: socdeg}
\textmd{SOCDEG} =
[\Degree(a_1 \in A), \ldots, \Degree(a_n \in A) ],
\end{equation}
where $n = \vert A \vert$ and the set $A$ is assumed to be ordered. In other words, the metric equals the degree of the CVE-labeled vertices in $G_s$. This degree centrality conveys a clear theoretical rationale. In the software engineering context this rationale relates to the saying ``too many cooks spoil the broth''. The essence behind the saying is that increasing number of participants increases the coordination requirements, which translate into delays in completing software engineering tasks~\cite{Bird06, Johnson16a}. There exists also some evidence for an assertion that bug resolution delays increase with increasing number of participants in the resolution processes \cite{Bhattacharya11}. Although the existing empirical evidence seems weak, the same dictum can be extended to a further hypothesis that ``too many developers'' increase the probability of introducing vulnerabilities during software development~\cite{Meneely10}. Given analogous reasoning, SOCDEG can be expected to lengthen the coordination delays. If a given $a \in A$ has a high degree, meaning that many participants posted messages containing the CVE, it may be that the vulnerability in question was particularly interesting or controversial. Either way, a longer delay could be expected for such a vulnerability.

In theory, also other vertex-specific centrality metrics could be used for modeling the delays. There are a couple of reasons to avoid additional centrality metrics, however. The first reason relates to interpretation ambiguities. For instance, the so-called closeness centrality is often used for quantifying information flows among a group of human participants \cite{BettenburgHassan13}. In the context of sender-receiver type of email networks \cite{Lubarski12, Tang14}, this quantification rests on the assumption that a reply to an email constitutes an information flow. In reality, however, the lack of a reply does not imply the absence of an information flow; a participant may read an email without replying~\cite{Nia10}. In addition to these theoretical limitations, the bipartite structure of $G_s$ makes the interpretation of many vertex centrality metrics challenging. For instance, the so-called betweenness centrality is often interpreted to reflect communication gatekeepers and information brokers~\cite{Bird06, PooCaamano17}, but it is difficult to theorize how a CVE identifier would be a gatekeeper. The second, more practical reason stems from multicollinearity issues induced by the inclusion of additional centrality metrics. As is typical~\cite{Ruohonen16MESSA, Schoch17}, many of the centrality metrics are correlated. In particular, SOCDEG is highly correlated with the betweenness centrality values.

Instead of explicitly computed centrality, the second social network metric takes a simpler approach to quantify the concept of core developers in the open source context. The definitions for such developers differ. For instance, some authors have identified core developers with a cutoff point for vertex degrees \cite{Licorish14, Toral10}, while others have relied on documents about developer responsibilities and commit accesses \cite{Conaldi13, Crowston17}. In the present context the relevant social network core is composed by the MITRE affiliates. Thus, MITREDEV is a dummy variable that scores one for all CVEs linked to the three labeled participants in Fig.~\ref{fig: core}.

The two remaining metrics are not explicitly related to social networks \textit{per se}, although both of these still proxy communication practices. Namely: the metric MSGSLEN counts the length of strings in all emails posted with a given CVE identifier divided by one hundred, while the metric MSGSENT records the Shannon entropy of these emails. Both metrics have been used previously in the software engineering context \cite{BettenburgHassan13}. The effect of both metrics upon $y_i$ can be also expected to be positive. Given the rationale of the mailing list for coordinating CVE identifiers, lengthy email exchanges and increasing entropy are both likely to increase the coordination delays. In other words, high values for these two metrics both run counter to the short ``use CVE-2016-6527''\cite{Openwall16c} communication patterns preferred on the list during the period observed.

\subsubsection{Infrastructure Metrics}\label{subsec: infrastructure metrics}

Six metrics are used for soliciting an answer to \ref{rq: infrastructure}. The first of these, INFDEG, is defined analogous to \eqref{eq: socdeg} but by using the vertex set $B$ present in the network $G_d$. Thus, this metric counts the number of semantically valid domain names in the adjacency of the CVE identifiers observed. The analytical meaning is similar to the number of hyperlinks in reports posted within bug tracking systems, which have been hypothesized to reflect bugs that are particularly difficult to remedy \citep{BettenburgHassan13}. By translating the same hypothesis to the vulnerability coordination context, the number of domain names extracted from the hyperlinks could be expected to increase the coordination delays. Accordingly, a CVE that accumulates many hyperlinks may correspond with a vulnerability that is particularly difficult to interpret. Another explanation may be that the signal of relevant information is lost to the noise of numerous hyperlinks. However, the effect of INFDEG could be alternatively speculated to shorten the delays. The rationale for this alternative speculation relates to the prerequisite constraints in typical software engineering coordination.

A requested CVE identifier is likely to end up in NVD in case it is backed by sufficient and valid technical information. In addition to INFDEG, these prerequisite constraints can be also retrospectively approximated through the references provided in NVD to the primary information sources. Like the historical \cite{Bozorgi10} and contemporary \cite{Ruohonen17TIR} databases, NVD maintains a list of reference sources that is frequently polled for new vulnerability information \cite{Christey13}. If a vulnerability appears in multiple sources, it is probable that a CVE for the vulnerability appears rapidly in NVD due to information gains about the technical details. Thus, the metric NVDREFS counts the number of NVD's reference sources for all CVEs observed. Even when a CVE was coordinated through \texttt{oss-security}, shorter delays can be expected for an identifier with multiple alternative sources for confirming the corresponding vulnerability.

\begin{table}[th!b]
\centering
\caption{Infrastructure Regular Expressions}
\label{tab: domain regex}
\begin{small}
\renewcommand*{\arraystretch}{1.2}
\begin{tabular}{ll}
\toprule
Metric & Regular expression for all $d \in D$ \\
\hline
VULNINF & \texttt{cert.}, \texttt{exploit-db.com}, \texttt{first.org}, \\
& \texttt{mitre.org}, \texttt{nist.gov}, \texttt{osvdb.org} \\
\cmidrule{2-2}
BUGS & \texttt{bugs.}, \texttt{bugzilla.}, \texttt{gnats.}, \texttt{issues.}, \\
& \texttt{jira.}, \texttt{redmine.}, \texttt{trac.}, \texttt{tracker.} \\
\cmidrule{2-2}
REPOS & \texttt{code.}, \texttt{cvs.}, \texttt{cvsweb.}, \texttt{download.}, \\
& \texttt{downloads.}, \texttt{ftp.}, \texttt{git.}, \texttt{gitweb.}, \\
& \texttt{hg.}, \texttt{packages.}, \texttt{svn.}, \texttt{webcvs.}, \texttt{websvn.} \\
\cmidrule{2-2}
SUPPORT & \texttt{blog.}, \texttt{blogs.}, \texttt{dev.}, \texttt{doc.}, \texttt{docs.},  \\
& \texttt{forum.}, \texttt{forums.}, \texttt{help.}, \texttt{info.}, \texttt{lists.}, \\
& \texttt{support.}, \texttt{wiki.} \\
\bottomrule
\end{tabular}
\end{small}
\end{table}

\begin{figure}[th!b]
\centering
\includegraphics[width=8cm, height=6cm]{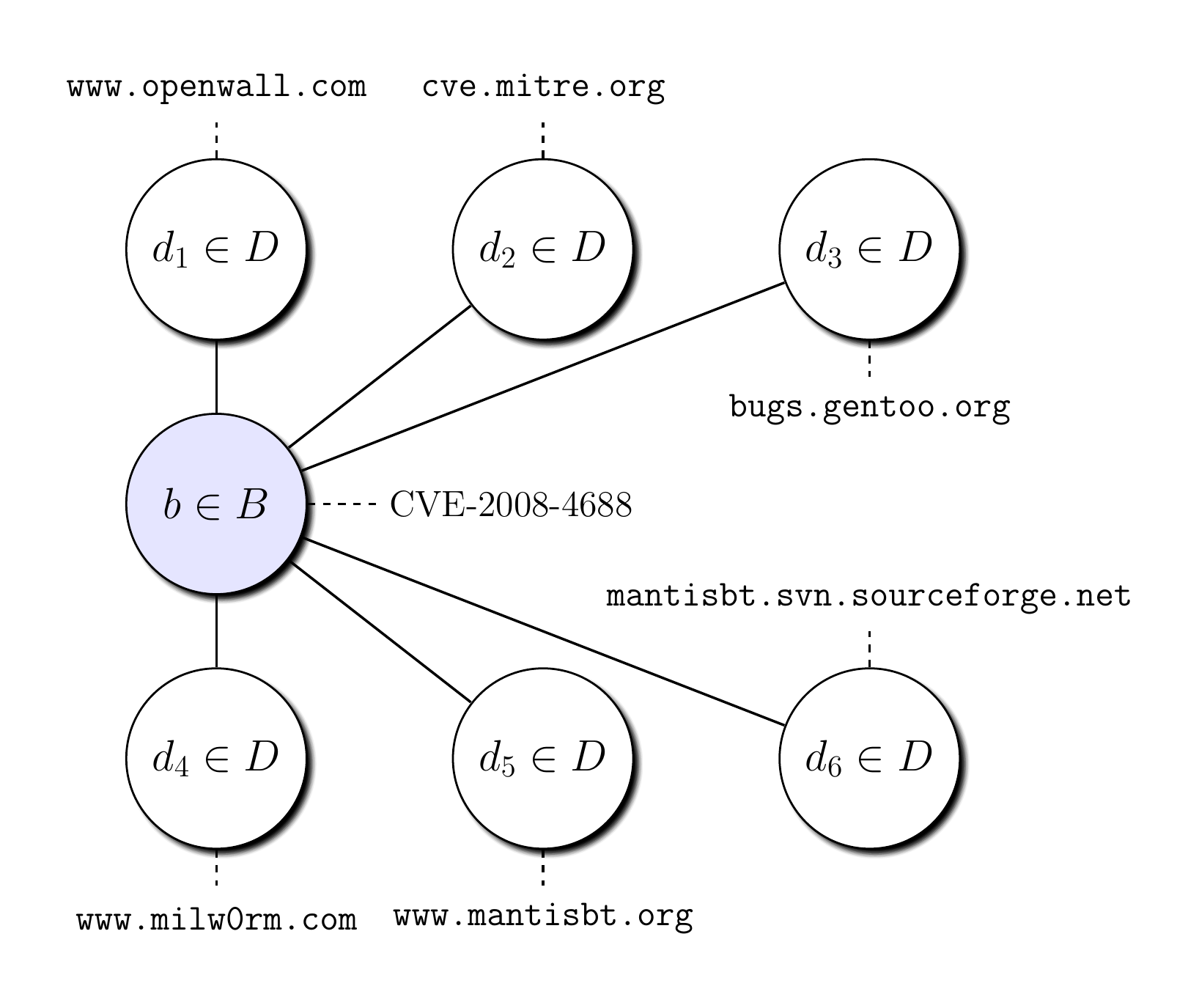}
\caption{An Example Domain Name Neighborhood}
\label{fig: example domain neighborhood}
\end{figure}

The same rationale can be used for justifying a more fine-grained look at the hyperlinks explicitly posted on the mailing list. The four remaining infrastructure metrics are dummy variables that approximate the content behind the domain names stored to the set $D$ in $G_d$. Given a domain name neighborhood of each CVE identifier in the vertex set $B$, the value of these metrics is zero unless at least one domain name in the neighborhood matches the regular expressions listed in Table~\ref{tab: domain regex}.

To further elaborate the operationalization of these metrics, Fig.~\ref{fig: example domain neighborhood} shows the neighborhood of a vertex labeled as CVE-2008-4688 in $B \in G_d$. For this CVE, the INFDEG metric takes a value six because there are six domain names linked to the identifier. Given the regular expressions, also the metrics VULNINF, BUGS, and REPOS each score a value one due to the vertices $d_2$, $d_3$, and $d_6$ (for the corresponding email see \cite{Openwall08b}). Of course, the expressions used are only approximations for automatically probing the nature of the primary tracking infrastructures. For instance, hosting services such as GitHub are mostly (but not entirely) excluded due to the matching primarily based on subdomains. Nevertheless, these infrastructure metrics convey a clear theoretical expectation. For instance, patches are often provided faster for high-profile vulnerabilities that affect multiple vendors, particularly in case computer emergency response teams are involved~\cite{Ruohonen16AICCSA, Temizkan12}. For analogous reasons, VULNINF could be expected to decrease the coordination delays observed. Given the open source way of coordinating defects in open bug trackers, also BUGS can be expected to show a significant negative impact upon the delays. In other words, a good bug report often goes a long way in satisfying a prerequisite constraint for a CVE assignment in the open source context.

\subsubsection{Vulnerability Metrics}\label{subsec: vulnerability metrics}

The remaining metrics are used for evaluating whether the severity and technical characteristics of the vulnerabilities coordinated affect the delays. Given the existing bug tracking research, there are good reasons to expect a correlation. Although the empirical evidence seems to be again somewhat mixed \cite{Bhattacharya11}, the severity of bugs have been observed to correlate with resolution delays in bug tracking systems \cite{ZhangHassan12, ZhouGupta15}. Furthermore, the complexity of source code has been observed to correlate with bug resolution times \cite{Kula10}. In terms of the metrics used, however, the bug tracking literature differs from vulnerability research.

Most bug tracking systems contain categories for assessing the severity and impact of bugs. The severity assignments within these systems are typically done either by bug reporters or by the associated developers. Due to the human work involved, the severity assignments have been observed to be relatively inconsistent and unreliable~\cite{TianHassan16}. In contrast to bug tracking systems, the severity of vulnerabilities archived to NVD are evaluated by experts by using the Common Vulnerability Scoring System (CVSS). Although it remains debatable whether the severity assignments are suitable for assessing the security risks involved~\cite{Allodi14, YonisMalaiya16}, the assignments themselves are highly consistent across different databases and evaluation teams~\cite{Johnson17}. Therefore, in technical terms,  CVSS provides a good and reliable framework for seeking an answer to the question \ref{rq: technical}.

The metrics in the models $\Model_4$ and $\Model_5$ are all based on the second version of the CVSS standard \cite{FIRST07}. (It is worth also remarking that only data based on this version is provided in NVD for the historical archival material relevant to the case studied.) This CVSS version classifies the severity of the vulnerabilities archived according to two dimensions: \textit{exploitability} possibilities and \textit{impact} upon successful exploitation. The impact dimension contains three metrics (confidentiality, integrity, and availability). Each of these can take a value from three options (\texttt{NONE}, \texttt{PARTIAL}, or \texttt{COMPLETE}). For instance, successfully exploiting the recently discovered and disclosed high-profile Meltdown vulnerability results in complete loss of confidentiality, although integrity and availability remain unaffected~\cite{NVD18a}. Due to multicollinearity issues, the three impact metrics IMPC, IMPI, and IMPA are based on collapsing the three value categories according to:

\begin{equation}\label{eq: cvss impact}
g(x) =
\begin{cases}
0 & \textmd{if} ~x~
\textmd{is}~\texttt{NONE}, \\
1 & \textmd{if}~x~\textmd{is}~
\texttt{PARTIAL}~\textmd{or}~\texttt{COMPLETE} .
\\
\end{cases}
\end{equation}

An analogous operationalization is used for the three dummy variables EXPNET, EXPCPLX, and EXPAUTH (see Table~\ref{tab: metrics}). These approximate the exploitability dimension based on the access \textit{vector} (whether exploitation requires a network or a local access), access \textit{complexity} (how complex are the access conditions for exploitation), and \textit{authentication} (whether exploitation requires prior authentication) metrics in the CVSS v.2~standard.

The CVSS standard and the associated data from NVD provide a wealth of information for approaching different security questions, but it is unclear how the re-coded impact and exploitability metrics may affect the coordination delays. There exists some evidence for a conjecture that the difficulty of reliable severity assessments vary in terms of what is being assessed and who is doing the assessments. This conjecture applies both to the CVSS framework \cite{Allodi17d} and to quantitative security assessments  in general \cite{Johnston18}. However, analogous reasoning does not seem to hold in the context of NVD and the second version of the CVSS standard; the content of the standard does not notably affect the delays for CVSS assignments~\cite{Ruohonen19ACI}. In addition to these empirical observations, it is difficult to speculate why some particular CVSS metric would either increase or decrease the coordination and related delays \cite{Ruohonen17COMSIS}. For these reasons, the CVSS metrics are used in the models without prior theoretical expectations. The assumption merely is that the severity of the vulnerabilities coordinated is statistically associated with the delays observed.

The final set of metrics is based on the Common Weakness Enumeration (CWE) framework maintained and developed by MITRE in cooperation with the associated governmental partners  and volunteers \cite{MITRE18a}. In essence, this comprehensive framework is used to catalog the typical ``root causes'' (weaknesses) that may lead to exploitable vulnerabilities in software products. The examples range from buffer and integer overflows to race conditions and software design flaws. By using a subset of frequent CWE identifiers \cite{MITRE18b}, NVD maintainers derive the underlying weaknesses behind the vulnerabilities archived either during the CVE coordination or in retrospect.

Currently, the CWE framework covers over 700 distinct weaknesses. Partially due to this large amount, some of the CWEs are difficult to evaluate in practice~\cite{GosevaPopstojanova15, WuGandhi10}. This difficulty does not necessarily imply that the framework itself would be problematic. Rather, many vulnerabilities chain a lot of distinct weaknesses together. For instance, CWEs can be used to exemplify that buffer overflow vulnerabilities posit a complex ``mental model'' for developers due to the tangled web of distinct programming mistakes involved~\cite{WuGandhi10}. Given analogous reasoning, it can be hypothesized that NVD-based CWE information can also explain a portion of the variation in the CVE coordination delays. As an example: when compared to XSS and CSRF vulnerabilities (as identified with CWE-79 and CWE-352), it may be that buffer overflow vulnerabilities (CWE-119) result in slower coordination because such vulnerabilities are usually technically more complex than simple web vulnerabilities. In other words, the effort required for interpretation may vary from a vulnerability to another, and such variation may show also in the coordination delays. To probe this general assumption, the final model $\Model_6$ includes the ten most frequent CWEs in the sample. Given that about 90\% of the CVEs in the sample have valid CWE entries in NVD, together these top-10 entries account for about 79\% of the CWEs available.

\subsection{Estimation}\label{subsec: estimation}

The coordination delay vector $\vcy = [y_1, \ldots, y_n]^\prime$ represent count data: the observations count the days between CVE assignments on the mailing list and the publication of the CVEs assigned within the primary global tracking database. Therefore, Poisson regression \cite{Ruohonen19ACI, Ruohonen17COMSIS} and regression estimators for survival analysis \cite{Ablon17, Temizkan12} have been typical statistical estimation strategies in the research domain. Previous work in comparable settings \cite{Arora10, daCosta18, Ruohonen19ACI, Ruohonen16AICCSA} indicates that the ordinary least squares (OLS) regression often works also sufficiently well when the dependent metric is passed through a transformation function $f(x) = \log(x + 1)$. This simple OLS approach is taken as the baseline for the empirical analysis. For all regression models estimated, the explanatory metrics with continuous scale are also transformed via the same function in order to lessen skew.

Although frequently used in applied research, the OLS approach contains also problems. In a sense, the transformation function is redundant and should be avoided in order to maintain the statistical properties of count data. It also adds small but unnecessary complexity for interpreting the parameter estimates. Furthermore, the residual vector $\vcepsilon$ from the OLS model is assumed to be independent and identically distributed from the normal distribution with a mean of zero and variance $\sigma^2$. This basic assumption can be written as $\E(\vcepsilon~\vert~\mtx_j) = \vczero$ and $\E(\vcepsilon\vcepsilon^\prime~\vert~\mtx_j) = \sigma^2\mti$, where $\E(\cdot)$ denotes the expected value, $\mtx_j$ the $j$:th model matrix for a given $\Model_j$ from Table~\ref{tab: metrics}, and $\mti$ an identity matrix. If the assumption is satisfied,
\begin{equation}\label{eq: ols}
\vcbeta~\vert~\mtx_j
\sim N(\vcbeta, \sigma^2[\mtx^\prime_j\mtx_j]^{-1}) ,
\end{equation}
where $\vcbeta$ is a regression coefficient vector and $N(\cdot)$ refers to the multivariate normal distribution. This distributional assumption allows exact inference based on $t$ and $F$ distributions. In many empirical software engineering applications the problem is not as much about the (asymptotic) normality assumption as it is about the unreliable variance patterns typically present in the typically messy datasets~\cite{Kitchenham17}. Thus, often, $\E(\vcepsilon\vcepsilon^\prime~\vert~\mtx_j) = \sigma^2\mtv$, where $\mtv$ is a generally unknown non-diagonal matrix that establishes some systematic pattern in the residuals. Such tendency is generally known as heteroskedasticity.

For instance, some vulnerability time series are known to exhibit time-dependent heteroskedastic patterns \cite{Ruohonen15COSE, Tang19}. In the present context heteroskedasticity is to be expected due to the count data characteristics, and possibly also due to the heavy use of dichotomous variables. Either way, it is important to emphasize that asymptotic inference is still possible under heteroskedasticity; the consistency of $\vcbeta$ remains unaffected but the estimates are inaccurate. An analogous consequence results from multicollinearity; the stronger the correlation between a variable and other variables, the higher the variance of the regression coefficient for the variable.

When heteroskedasticity is an issue, more accurate statistical significance testing can be done by adjusting the variance-covariance matrix, $\Var(\vcbeta~\vert~\mtx_j) = \sigma^2[\mtx^\prime_j\mtx_j]^{-1}$, with well-known techniques based on the unknown $\mtv$ estimated from data. The OLS results reported use the MacKinnon-White adjustment \cite{MacKinnonWhite85} conveniently available from existing implementations \cite{Zeileis04}. Another point is that rather analogous problems are often encountered with count data regressions. For instance, the negative binomial distribution is often preferable over the Poisson distribution \cite{Ruohonen19ACI}. Also the gamma distribution is known to characterize related difference-based vulnerability datasets~\cite{Johnson16b}. Furthermore, conventional methods for survival analysis face some typical problems. Although not worth explicitly reporting, it can be shown that the Cox regression's so-called proportional hazards assumption fails for most of the metrics in $\Model_1, \ldots, \Model_6$, for instance. Due to these and other reasons, it beneficial to use a regression estimator that makes no distributional assumptions. Quantile regression (QR) is one of such estimators.

Ordinary least squares provides estimates for the conditional mean. Quantile regression estimates quantiles. The $\tau$:th quantile is defined by
\begin{equation}
x_\tau = \inf\lbrace x~\vert~F(x) \geq \tau \rbrace ,
\quad 0 \leq \tau \leq 1,
\end{equation}
where $F(\cdot)$ denotes a cumulative distribution function, while the \change{infimum} operator is used to denote the smallest real number for which the condition $F(x) \geq \tau$ is satisfied. If $\tau = 0.5$, for instance, QR provides a solution $\hat{\vcbeta}$ for the conditional median of $\vcy$ conditional on $\mtx_j$. Estimation minimizes the sum of absolute residuals:
\begin{equation}\label{eq: qr}
\min_{\vcbeta}\sum^n_{i=1}
\rho_\tau (y_i - \beta_0 - \vcx^\prime_{i(j)}\vcbeta) ,
\end{equation}
where $\beta_0$ is the intercept, $\vcx^\prime_{i(j)}$ is the $i$:th row vector from the $j$:th model matrix $\mtx_j$, and $\rho_\tau(x) = x[\tau - I(x < 0)]$ with $0 < \tau < 1$ and $I(\cdot)$ denoting an indicator function~\cite{Alhamzawi12, Koenker99}. The optimization of \eqref{eq: qr} is computed with a so-called Frisch-Newton algorithm available in existing implementations \cite{quantreg18}. The benefits from QR are well-known and well-documented. Among these are the lack of distributional assumptions and the robustness against outliers. Although asymptotic assumptions are still required for significance testing under heteroskedasticity, no parametric assumptions are made regarding the residual vector~\cite{Koenker82}. For applied research, quantile regression has also an immediate appeal in that the whole range in the conditional distribution of the explained variable can be observed. This potential is also relevant for vulnerability delay metrics, which typically (but not necessarily) tend to be distributed from a long-tailed distribution.

There are a couple of additional points that still warrant brief attention. First, potential non-linearities should be accounted for also with QR. As the dataset contains many variables but only five of these have a $\Continuous$ scale (see Table~\ref{tab: metrics}), non-linear modeling of the explanatory metrics can be reasonably left for further work, however. The $\log(x + 1)$ transformation applied to these five variables also lessens some of these concerns (particularly regarding MSGSLEN and MSGSENT). Second, multicollinearity is always a potential issue with typical empirical software engineering datasets. Although the concern is not as pressing as with software source code metrics \cite{Fenton99}, analogous datasets containing social network and communication metrics allow to also expect potential multicollinearity issues \cite{BettenburgHassan13}. Instead of adopting techniques such as principal component regression---which would make the interpretation of Fig.~\ref{fig: hypotheses} difficult---the QR models are re-checked with the least absolute shrinkage and selection operator (LASSO). For quantile regression, LASSO amounts to optimizing
\begin{equation}\label{eq: qr lasso}
\min_{\beta_0, \vcbeta}\sum^n_{i=1}
\rho_\tau( y_i - \beta_0 - \vcx^\prime_{i(j)}\vcbeta) +
\lambda\lVert \vcbeta \rVert_1 ,
\end{equation}
where $\lVert \cdot \rVert_1$ denotes the $L_1$-norm and $\lambda$ is a non-negative tuning parameter \cite{Alhamzawi12}. Although cross-validation and other techniques can be used for selecting the tuning parameter \cite{Ruohonen19ACI}, the QR-LASSO regressions are estimated with $\lambda \in [1, 2, \ldots, 100]$ using the full $\Model_6$. This range captures most of the regularization applicable to the dataset.

The rationale behind \eqref{eq: qr lasso} follows the rationale of LASSO in general: when $\lambda$ increases, the quantile regression coefficients shrink toward zero. This regularization makes LASSO useful as a variable selection tool for high-dimensional and ill-conditioned datasets. When $\lambda$ is sufficiently large, some of the coefficients shrink to zero, leaving a group of more relevant coefficients. The model selection properties are not entirely ideal, however. When a coefficient for a correlated variable is regularized to zero, the coefficients of the other correlated variables are affected; that is, LASSO tends to select one correlated variable from a group of highly correlated variables \cite{LeeChen17}. It is worth to remark that also the common alternatives contain problems. In addition to issues related to multiple comparisons, the expected heteroskedasticity presumably causes problems for a stepwise variable selection algorithm particularly in case the algorithm uses statistical significance to make decisions. Instead of relying on such algorithms, a more traditional is adopted for the modeling and inference.

\subsection{Modeling}

Applied regression modeling has two essential functions. It can be used to make predictions based on explanatory metrics, or it can be used to examine ``the strength of a theoretical relationship'' between the explained metric and the explanatory metrics \cite{Giacalone17}. Already because prediction is not a sensible research approach for a historical case study, this paper leans toward the examination of theoretical relationships based on classical statistical inference. The two functions cannot be arguably separated from each other, however. Given the notorious data limitations affecting vulnerability archiving \cite{Christey13}, the inevitable noise introduced by the collection of online data, and many related reasons, a watchful eye should be kept for asserting the presence of a signal in terms of prediction. The magnitudes of the regression coefficients, the effect sizes, are used to balance the final subjective judgment calls regarding the research questions \ref{rq: social}, \ref{rq: infrastructure}, and \ref{rq: technical}.

The six models $\Model_1, \ldots, \Model_6$ are fitted consecutively. This hierarchical modeling approach \cite{BettenburgHassan13} is also known as a bottom-up or specific-to-general modeling strategy \citep{Lutkepohl07}. For comparing the six OLS models, the adjusted coefficients of determination (adj.~$R^2$) and Akaike's information criterion (AIC) values are used. Higher and lower values, respectively, indicate better performance according to these two common evaluation statistics. Both penalize the performance by the number of parameters. Although pseudo-$R^2$ measures are available for quantile regression~\cite{Koenker99}, AIC is used to summarize the general signal-to-noise-type of performance of four quantile regressions estimated for each of the six models. The following quantiles are used in order to have probes across a wide range of the conditional distribution of the coordination delays:
\begin{equation}\label{eq: taus}
\vctau = [ 0.25, 0.50, 0.75, 0.90 ] .
\end{equation}

It should be noted that neither censoring nor transformations are applied for the delays when estimated with QR. Consequently, the predicted values may be also negative, which is theoretically impossible. As prediction is not a goal, this limitation can be accepted.

The specific-to-general strategy is used also for formal testing with two analytical dimensions. First, the nested structure is exploited to test parameter restrictions between models. The testing is done by comparing (restricted) $\Model_{j-1}$ against (unrestricted) $\Model_j$ for $1 < j \leq 6$. The procedure follows the standard (Wald's) logic to compare nested models \cite{Koenker99}. In fact, also the test statistics are delivered via $F$-like statistics \cite{Koenker82}. Second, QR provides also means to test whether the coefficients are equal for a given $\Model_j$ across the whole \eqref{eq: taus} or subsets thereof. For instance, to evaluate whether the effect of WEEKEND remains constant across the proxied conditional range of the coordination delays, the twentieth coefficients from $\lbrace \hat{\vcbeta}_{\tau_1}, \hat{\vcbeta}_{\tau_2}, \hat{\vcbeta}_{\tau_3}, \hat{\vcbeta}_{\tau_4}\rbrace_j$ for the $j$:th model would be used to test that $\hat{\beta}_{20j\tau_1}
\simeq \hat{\beta}_{20j\tau_2}
\simeq \hat{\beta}_{20j\tau_3}
\simeq \hat{\beta}_{20j\tau_4}$. As it is reasonable to speculate that the effect of the explanatory metrics differ particularly at the tails, the coefficients in the following sets are used for the between-quantile testing:
\begin{align}\label{eq: quantile sets}
S_1 &= \lbrace\
\hat{\vcbeta}_{\tau_1}, \hat{\vcbeta}_{\tau_2}
\rbrace_j,
\\ \notag
S_2 &= \lbrace\
\hat{\vcbeta}_{\tau_1}, \hat{\vcbeta}_{\tau_2}, \hat{\vcbeta}_{\tau_3}
\rbrace_j,
\\ \notag
S_3 &= \lbrace\
\hat{\vcbeta}_{\tau_1}, \hat{\vcbeta}_{\tau_2}, \hat{\vcbeta}_{\tau_3}, \hat{\vcbeta}_{\tau_4}
\rbrace_j .
\end{align}

The bootstrap procedure available from the implementation used~\cite{quantreg18} is used to compute the statistical significance for the between-model tests. To accompany the MacKinnon-White adjustment for the OLS estimates, the same bootstrap procedure is used also for reporting the statistical significance of the coefficients from the QR regressions. Unfortunately, analogous procedure has not been implemented for the between-quantile tests. The conventional approach \cite{Koenker82} is thus used instead.

Finally, the regression analysis is accompanied with a brief classification experiment to further assess the general performance. Following the existing bug tracking research~\cite{Habayeb18, ZhangHassan12}, this experiment is conducted by splitting the coordination delays into low-delay and high-delay groups according to median. Although this sample splitting is a good example of data manipulation that can be avoided with quantile regression \cite{Koenker01}, it provides a simple additional assertion regarding the overall performance across the six models. Classification accuracy is a sufficient evaluation metric for this simple purpose. Estimation is done with a readily available and well-known random forest classifier~\cite{caret, randomForest}. Ten-fold cross-validation is used during training. Given that prediction of new data is not a realistic scenario for the historical \texttt{oss-security} case studied, testing is done with a randomly picked set containing ten percent of the CVEs observed. The same test set is used for all six classification models.

\section{Results}\label{section: results}

The results are disseminated by first presenting a few relevant descriptive statistics. A summary of the regression analysis follows. In addition, four computational checks are presented about the statistical performance.

\subsection{Descriptive Statistics}

\begin{figure}[th!b]
\centering
\includegraphics[width=\linewidth, height=5cm]{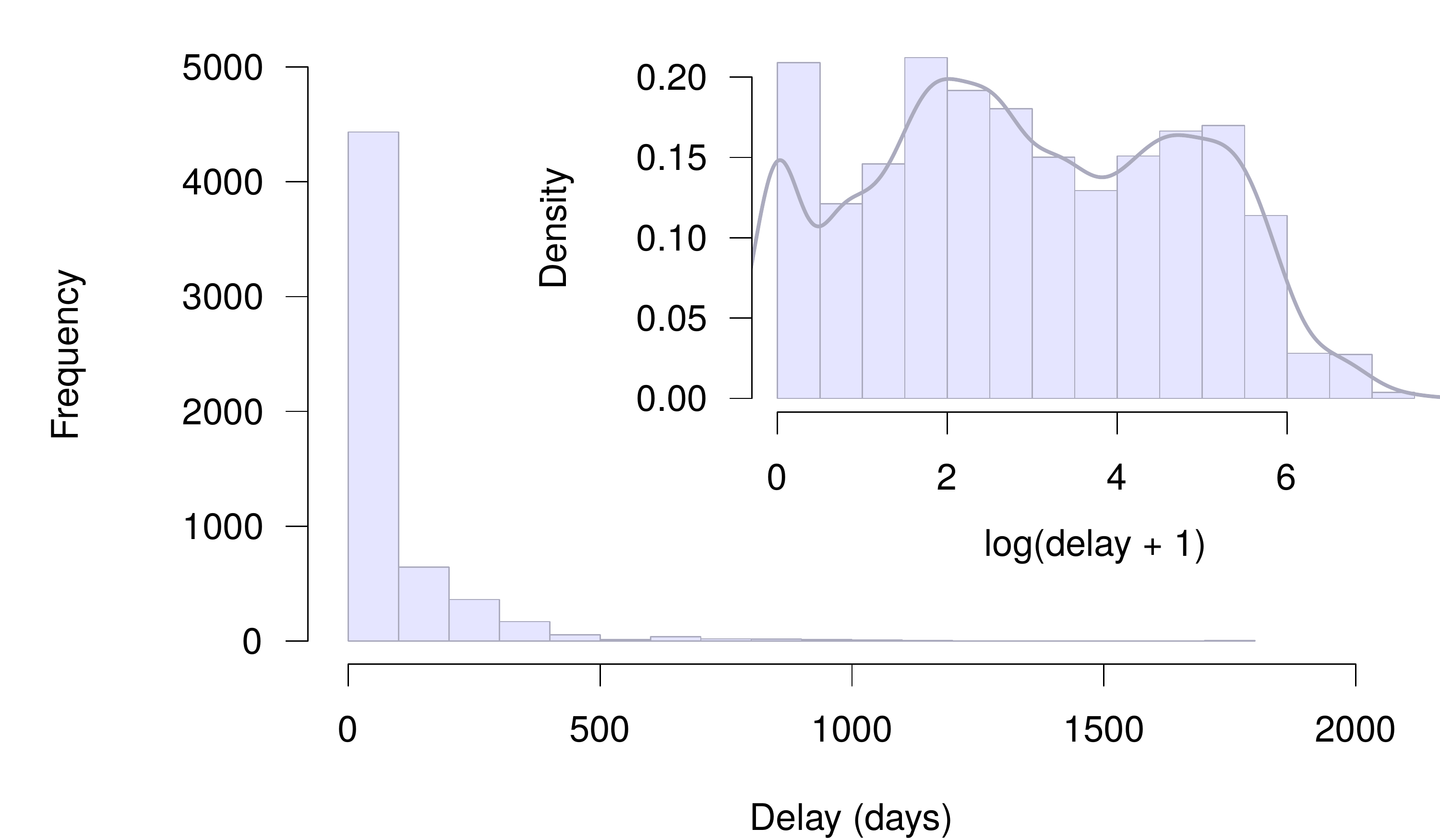}
\caption{Coordination Delays}
\label{fig: delays}
\end{figure}

The sample contains $n = 5,780$ identifiers once the exclusion criteria in \eqref{eq: delays restriction} is enforced (see Table~\ref{tab: sample}). These were discussed by about five hundred unique participants who posted hyperlinks containing $4,642$ unique domains. The mean and median delays were $77$ and $15$ days, respectively. The values for \eqref{eq: taus} are $4$, $15$, $92$, and $226$ days. As can be seen from the histogram in the outer plot of Fig.~\ref{fig: delays}, the delay distribution indeed has a long tail. This tail contributes to the large standard deviation of $147$ days. Thus, there is a large majority group of CVEs that were coordinated rapidly and a small but important group of CVEs for which the coordination was significantly delayed. For a few outlying CVEs, the coordination has taken even over four years. To balance this remark, it can be noted that about 10\% of the cases observed attain a value zero, meaning that these CVEs appeared in NVD during the same day when these were requested on the mailing list.

\begin{table}[th!b]
\centering
\caption{Sample Characteristics}
\label{tab: sample}
\begin{small}
\renewcommand*{\arraystretch}{1.2}
\begin{tabular}{lr}
\toprule
Quantity & Value \\
\hline
$\vert A \vert = $ Number of CVEs & 5,780 \\
$\vert P \vert = $ Number of participants & 496 \\
$\vert D \vert = $ Number of domain names & 4,642 \\
\bottomrule
\end{tabular}
\end{small}
\end{table}

\begin{figure}[th!b]
\centering
\includegraphics[width=\linewidth, height=7cm]{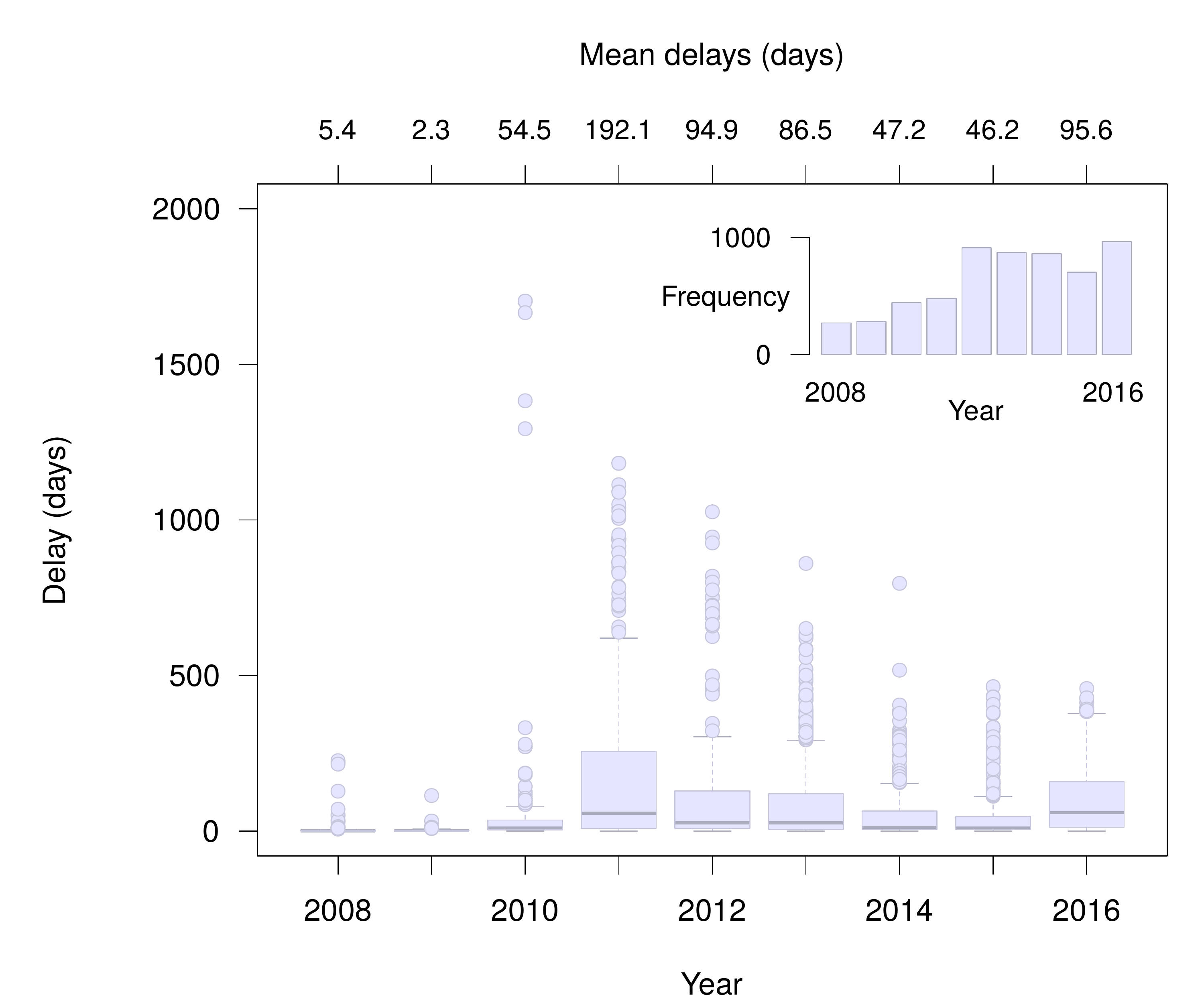}
\caption{Annual Coordination Delays}
\label{fig: delays annual}
\end{figure}

\begin{figure*}[t!hb]
\centering
\includegraphics[width=18cm, height=14cm]{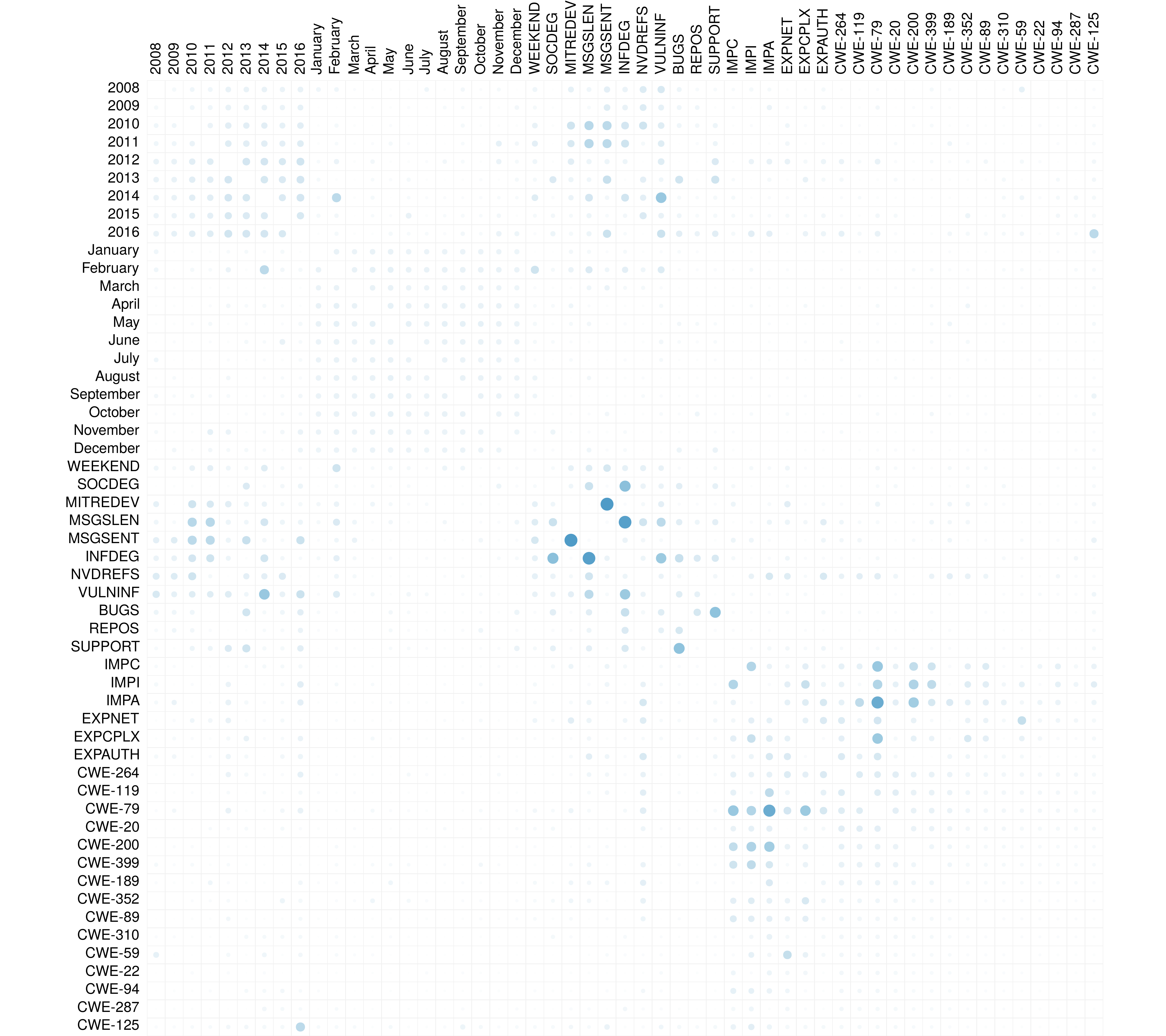}
\caption{Correlations Between Explanatory Metrics (absolute values of Spearman's rank correlation coefficients)}
\label{fig: cor}
\end{figure*}

As can be seen from the inner plot in Fig.~\ref{fig: delays}, the transformation function $f(x) = \log(x + 1)$ does not yield normally distributed delays, although the shape of the delay distribution is still better suited for OLS regression after the transformation. To examine whether immediate multicollinearity issues are also visible, Fig.~\ref{fig: cor} displays the absolute correlation coefficients between all explanatory variables. (As with the regression analysis, the transformation function is used also for the continuous variables marked with the symbol $\Continuous$ in Table~\ref{tab: metrics}.) The interpretation is simple: the bigger a circle and the darker a color, the stronger the correlation between any two metrics. Given this simple visual decoding guide, there are two notable correlations that warrant a multicollinearity concern.

The first is between MITREDEV and MSGSENT. The correlation between these two metrics reflects the typical ``use this identifier'' replies made by the MITRE affiliates. In other words, such replies tend to result in a lower entropy for the messages referencing CVEs. The second notable correlation is between INFDEG and MSGSLEN. Also this correlation is logical: posting many hyperlinks increases the length of the messages. Despite of these correlations, the regression estimates do not change notably when only MITREDEV and INFDEG (or, equivalently, MSGSENT and MSGSLEN) are included in the models. The same applies to the few visible correlations between the CVSS and CWE metrics. For these reasons, the results reported in the subsequent section are based on the original model specifications summarized in Table~\ref{tab: metrics}.

To examine the longitudinal variation, the outer plot in Fig.~\ref{fig: delays annual} shows the delays across the period observed. This illustration clearly shows that the delays started to increase from 2010 onward, but the increases were mostly caused by outlying CVEs. The arithmetic mean delays shown on the upper $x$-axis indicate that the average annual delays have been below one hundred days. The annual median delays are all below $60$ days. These averages align roughly with the so-called \textit{grace periods} (the time vendors are given to patch their products before public release of information) typically used in the security industry during vulnerability disclosure. Although the lengths of these grace periods vary, an upper limit of about three months seems to capture most explicitly reinforced policies \cite{McQueen09, Ruohonen16AICCSA}. Finally, the inner plot in Fig.~\ref{fig: delays annual} shows that the increasing delays and the increasing variation corresponded with the increasing amount of CVEs coordinated. The increased coordination volume in turn corresponded with the increased number of participants~\cite{Ruohonen17IWSMMensura}. These longitudinal changes allow to expect that the baseline model $\Model_1$ explains a relatively large share of the total variation in the delays. This expectation provides a good way to start the dissemination of the results from the regression analysis.

\subsection{Model Performance}

The statistical performance is somewhat modest regardless of the model. Analogous to interpreting effect sizes~\cite{Kampenes08}, adjectives such as modest are subject to interpretation, of course. Values $R^2 < 0.3$ are neither atypical in the vulnerability research domain~\cite{Arora10} nor uncommon in empirical software engineering experiments in general. Such values are commonly seen also in social sciences, including economics \cite{Karlsson19}. In this sense, the about 28\% of the total variation explained by the OLS model $\Model_6$ indicates modest but typical performance for regressions involving human beings. This observation can be seen from Table~\ref{tab: performance}, which shows the adjusted coefficients of determination, the AIC values, and the differences between the AIC values of the consecutively estimated models. The latter two are shown also for each of the four per-model QR regressions.

\begin{table*}[th!b]
\centering
\caption{Model Performance}
\label{tab: performance}
\begin{small}
\renewcommand*{\arraystretch}{1.2}
\begin{tabular}{llrrrcrrrrrrrr}
\toprule
&& \multicolumn{3}{c}{OLS}
&& \multicolumn{8}{c}{QR} \\
\cmidrule{3-5}\cmidrule{7-14}
&& \multicolumn{3}{c}{(mean)} &&
\multicolumn{2}{c}{$\tau = 0.25$} &
\multicolumn{2}{c}{$\tau = 0.50$} &
\multicolumn{2}{c}{$\tau = 0.75$} &
\multicolumn{2}{c}{$\tau = 0.90$} \\
$\Model_k$ && Adj.~$R^2$ & AIC & $\Delta$AIC && AIC & $\Delta$AIC & AIC & $\Delta$AIC & AIC & $\Delta$AIC & AIC & $\Delta$AIC \\
\cmidrule{1-1}\cmidrule{3-5}\cmidrule{7-14}
$\Model_1$ && 0.240 & 21841 &  --
&& 64775 & -- & 68490 & -- & 73136 & -- & 77920 & -- \\
$\Model_2$ && 0.250 & 21774 & -67
&& 64772 & -3 & 68472 & -18 & 72956 & -180 & 77032 & -888 \\
$\Model_3$ && 0.276 & 21572 & -202
&& 64771 & -1 & 68428 & -44 & 72757 & -199 & 76561 & -471 \\
$\Model_4$ && 0.278 & 21563 & -9
&& 64776 & +5 & 68426 & -2 & 72740 & -17 & 76544 & -17 \\
$\Model_5$ && 0.277 & 21567 & +4
&& 64781 & +5 & 68431 & +5 & 72741 & +1 & 76545 & +1 \\
$\Model_6$ && 0.281 & 21552 & -15
&& 64796 & +15 & 68433 & +2 & 72721 & -20 & 76504 & -41 \\
\bottomrule
\end{tabular}
\end{small}
\end{table*}

Four further points can be made about the performance. First, as expected, the longitudinal control variables provide most of the explanatory power. When compared to $\Model_1$, the social network and communication metrics increase the performance by about one percentage point according to the OLS estimates. The infrastructure metrics subsequently increase the performance by about three percentage points according to the adjusted $R^2$ values. Second, the conditional median ($\tau = 0.5$) and the conditional log-transformed mean (OLS) models indicate comparable behavior in terms of $\Delta$AIC. Third, the QR estimates indicate that the longer the delays, the bigger the performance gains from $\Model_2$ and $\Model_3$. Particularly when the tail is probed with $\tau = 0.9$, the social network, communication, and infrastructure metrics reduce the AIC values substantially. In contrast, when short delays are proxied with $\tau =  0.25$, these metrics do not bring performance gains. Last: at best, there are only small improvements after $\Model_3$, and there are also cells with positive $\Delta$AIC increments.

\begin{figure}[th!b]
\centering
\includegraphics[width=\linewidth, height=7cm]{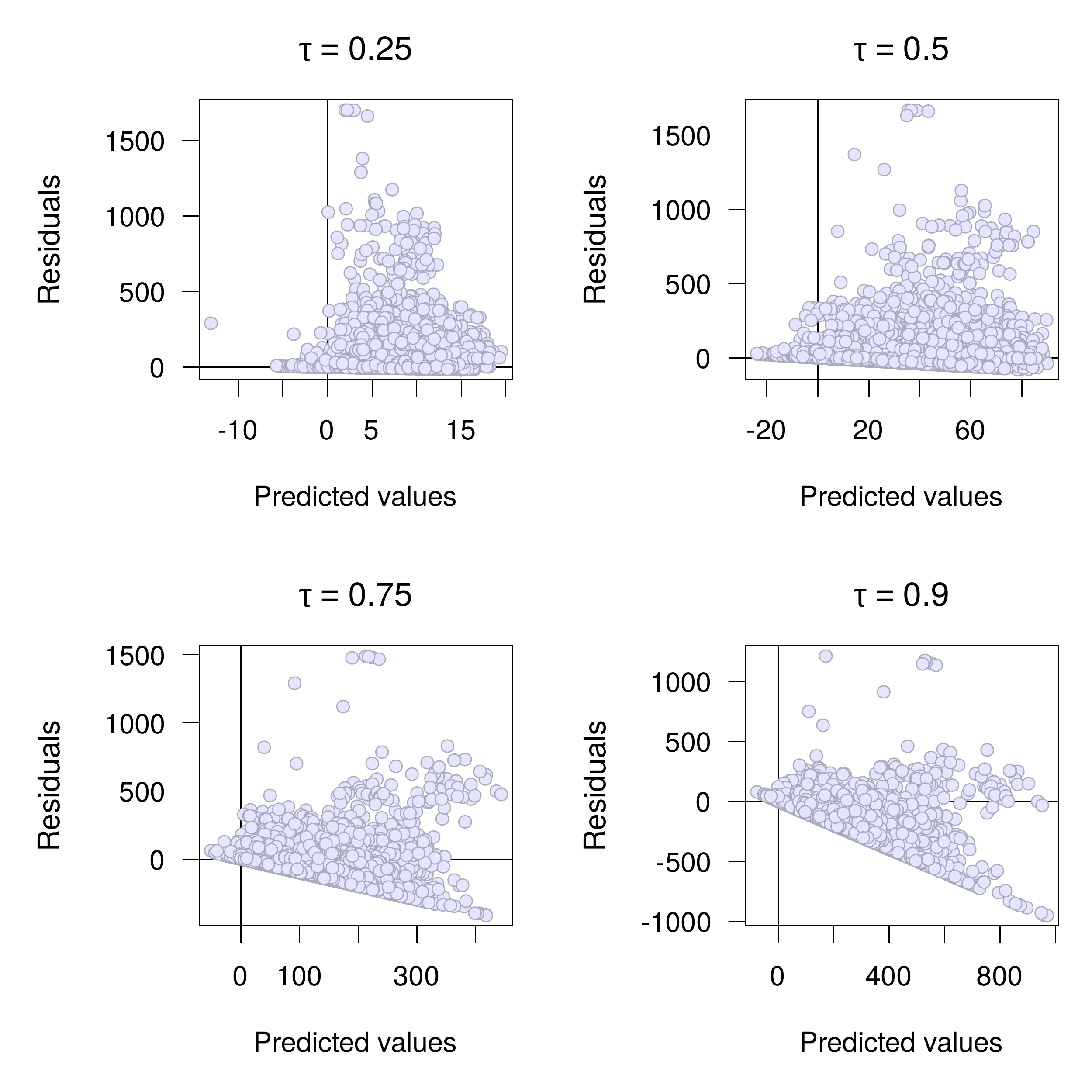}
\caption{Residuals and Fitted Values (QR, $\Model_6$)}
\label{fig: qr residuals}
\end{figure}

However, all but one of the consecutively estimated nested models yield statistically significant rejections of the between-model parameter restrictions. In other words, 95\% of the hypotheses that the smaller (restricted) models would be adequate compared to the larger models are rejected according to the bootstrapped computations. It may also be that information criteria measures such as AIC have problems in differentiating between models when the overall performance is modest~\cite{Karlsson19}. The turn to statistical significance motivates to also take a look at the residuals from the full $\Model_6$ in the form of Fig.~\ref{fig: qr residuals}. The residuals are rather randomly scattered across the $x$-axes only for the conditional median regression. For $\tau = 0.25$, there is a $\cap$-shaped pattern. For $\tau = 0.75$ and particularly for $\tau = 0.9$, there exists a more visible linear heteroskedasticity pattern; the longer the delays estimated, the larger the residuals. It is beyond the scope of this paper to review and evaluate how the QR-based inference performs under the heteroskedasticity patterns observed. Nevertheless, it seems reasonable to prefer the conditional median regressions and proceed with caution particularly when interpreting the $\tau > 0.5$ quantile regressions.

\subsection{Regression Estimates}

\begin{figure}[p!]
\centering
\includegraphics[width=\linewidth, height=22cm]{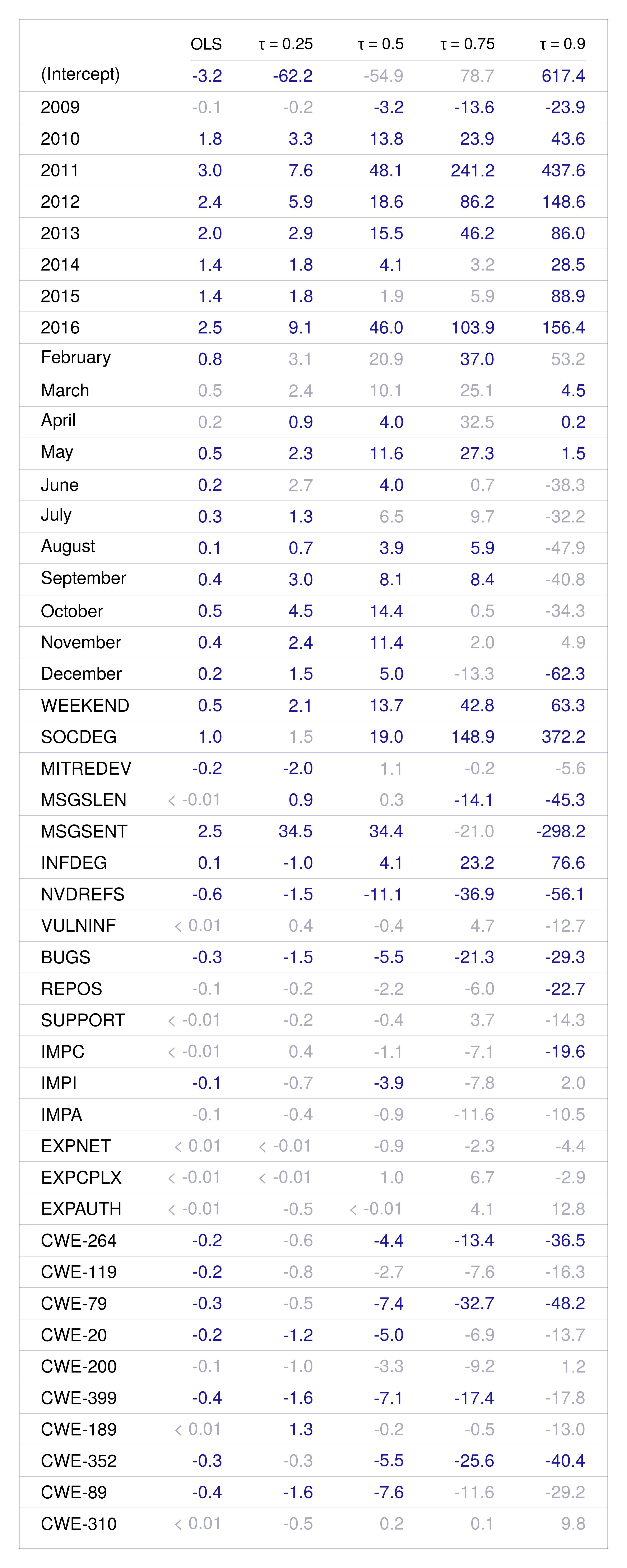}
\caption{Regression Coefficients ($\Model_6$, darker blue values denote $p < 0.05$, MacKinnon-White standard errors for the OLS regression and bootstrapped standard errors for the quantile regressions)}
\label{fig: coefficients}
\end{figure}

\begin{figure}[p!]
\centering
\includegraphics[width=\linewidth, height=22cm]{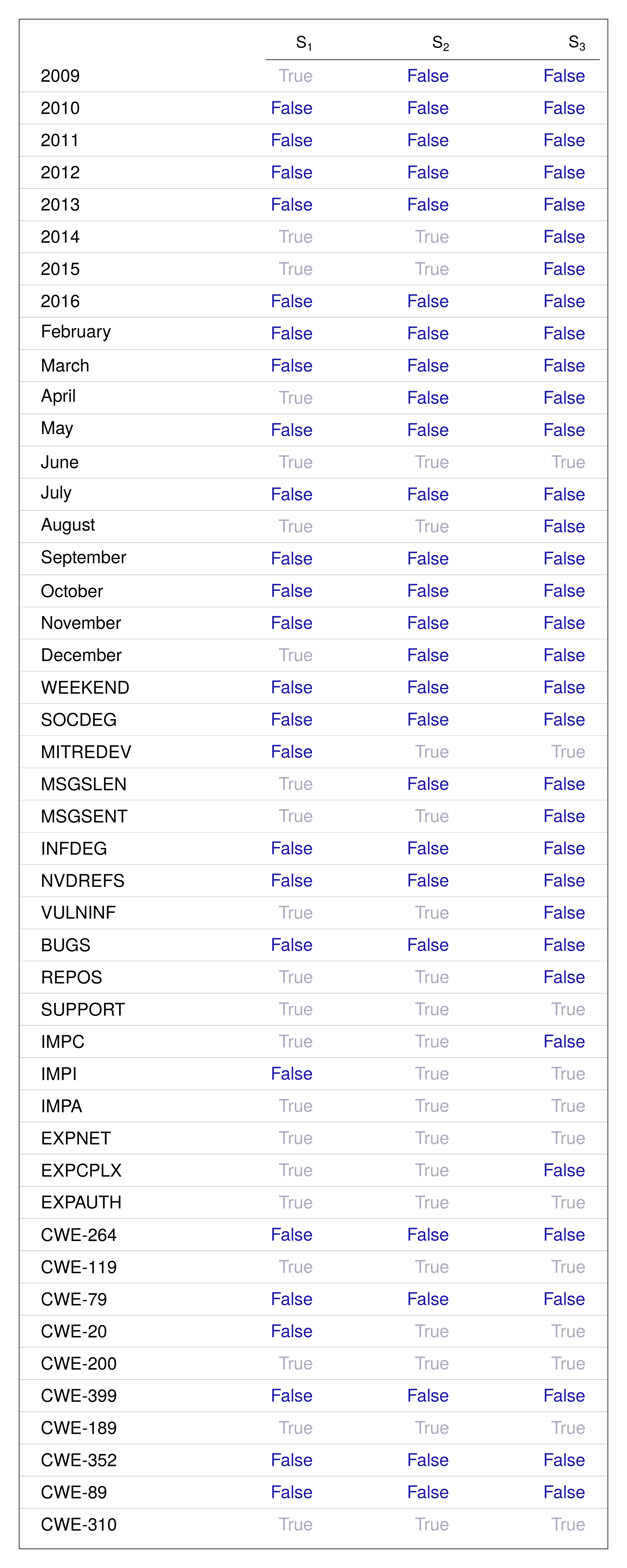}
\caption{Between-Quantile Tests (given the full unrestricted $\Model_6$, the cells show whether $p \geq 0.05$ for testing the null hypotheses that the slope coefficients are equal with respect to the sets in Eq.~\ref{eq: quantile sets})}
\label{fig: between-quantile tests}
\end{figure}

The regression estimates from the full unrestricted model are shown in Fig~\ref{fig: coefficients}. The accompanying Fig.~\ref{fig: between-quantile tests} shows the results from the between-quantile tests. As noted in the previous section, these tests as well as the statistical significance of the coefficients should be interpreted tentatively. While keeping this point in mind, the following enumeration summarizes the key observations.

\begin{itemize}
\item{The effects of the longitudinal control metrics are consistently strong. When compared to 2008, the increased delays particularly in 2011, 2012, and 2016 are visible especially with respect to the $\tau = 0.9$ quantile regression. The same information is conveyed in Fig.~\ref{fig: delays annual}. There exists also monthly variation in the delays. In addition to annual holidays, another reason may relate to security conferences and related events that tend to spike the public disclosure of new vulnerabilities~\cite{JohMalaiya17}. In any case, it is simpler to interpret the positive effect of WEEKEND. When compared to other days of the week, the median delays have been about two weeks longer for requests made on weekends. The effect of WEEKEND also increases at the tails of the conditional delay distribution.}
\item{The social network and communication metrics exhibit strong effects. In particular, the assumption about ``too many cooks'' seem to hold well; the effect of $\log(\textmd{SOCDEG} + 1)$ is large particularly for long coordination delays. Also the communication surrogates $\log(\textmd{MSGSLEN} + 1)$ and  $\log(\textmd{MSGSENT} + 1)$ show large effects. The signs vary, however. Because the coefficients are positive for $\tau \leq 0.5$ and negative for $\tau \geq 0.75$, the prior theorization in Subsection \ref{subsec: social network and communication metrics} should not be as unequivocal as was presented.}
\item{From the infrastructure metrics, the coefficients for INFDEG, NVDREFS, and BUGS are statistically significant for each regression. The coefficient magnitudes are also notable. If a CVE request was accompanied with a hyperlink to a bug tracking system, the median delay was about five to six days shorter, for instance. As was expected (see Subsection~\ref{subsec: infrastructure metrics}), the signs are also negative for NVDREFS and BUGS but positive for INFDEG. It seems that hyperlinks can also increase the noise, which tends to increase the CVE coordination delays.}
\item{The CVSS and CWE metrics show diverging results. On one hand, only three of the coefficients for the CVSS metrics are statistically significant in the five $\Model_6$ regressions. The effect sizes are also small, and mostly equal according to the between-quantile tests. On the other hand, many of the CWE metrics attain statistically significant coefficients with large magnitudes. All of the CWE metrics that are statistically significant have negative signs. As these observations apply also for the predominantly web-related weaknesses (CWE-79, CWE-89, and CWE-352; to some extent, also CWE-20), it seems that mundane low-profile vulnerabilities are generally coordinated faster. As was discussed in Subsection~\ref{subsec: vulnerability metrics}, interpretation is not easy, however. The difficulty of interpretation further increases because some of the CWE metrics likely proxy the effects of the CVSS metrics due to multicollinearity (see Fig.~\ref{fig: cor}). All in all, it can be concluded that the coordination delays vary also in terms of the technical characteristics of the vulnerabilities coordinated. As will be shown in the next section, performance can be also slightly increased with additional weaknesses, although the signals from the CVSS and CWE metrics still remain somewhat weak.}
\end{itemize}

\subsection{Computational Checks}

\begin{figure}[th!b]
\centering
\includegraphics[width=\linewidth, height=9cm]{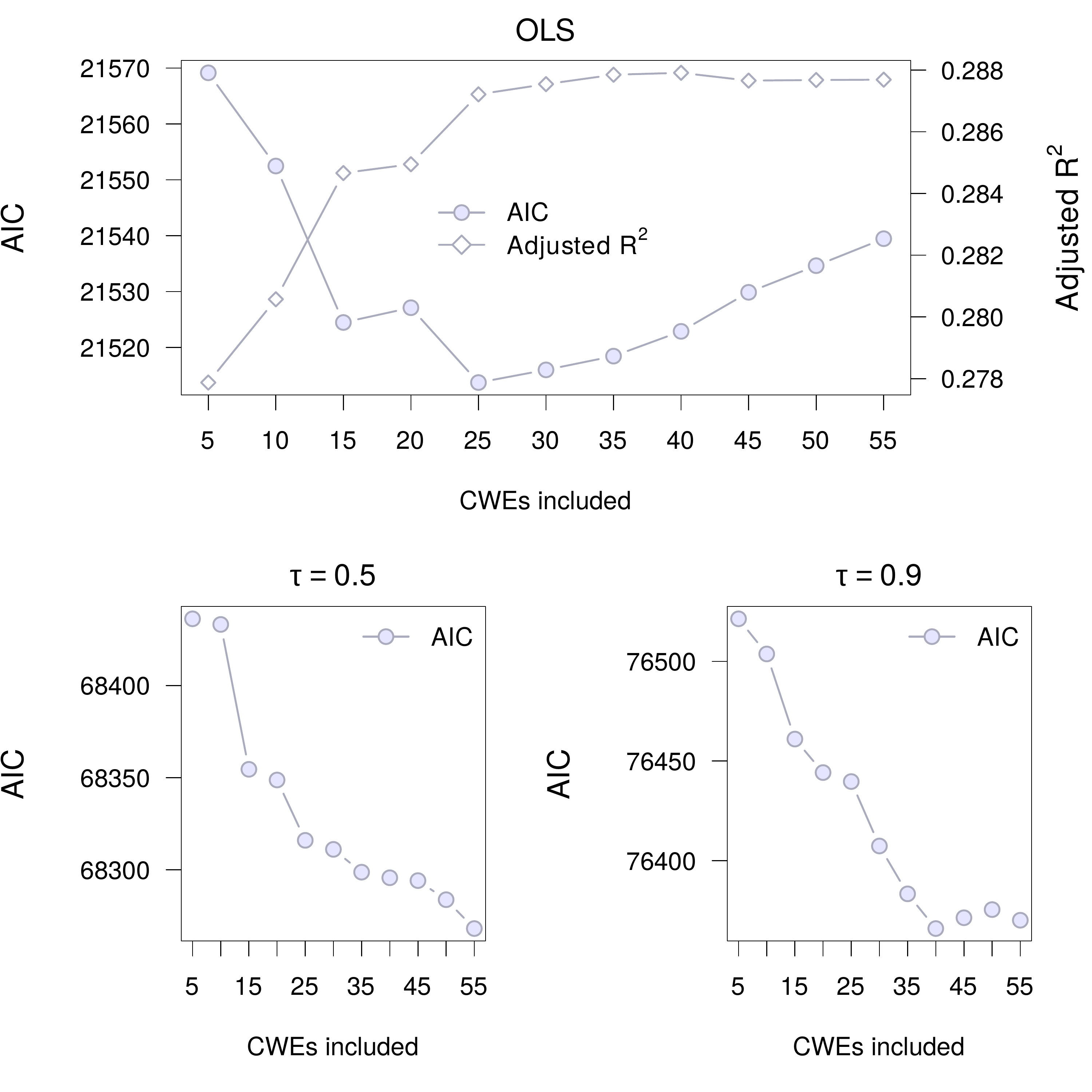}
\caption{Model Performance with Varying Number of CWEs}
\label{fig: cwe performance}
\end{figure}

Four computational checks are made to assess the robustness of the conclusions. The first check relates to the number of CWEs included; the model $\Model_6$ includes the ten most frequent CWEs in the sample, but there are $55$ unique CWEs in total. The summary shown in Fig.~\ref{fig: cwe performance} displays the performance when separate models are estimated by adding $5, 10, 15, \ldots, 55$ weaknesses to the fifth model specification. The case with ten weaknesses thus equals $\Model_6$. The OLS performance increases up to around $25$ weaknesses. For the median QR regression, all of the fifty-five CWEs seem to steadily reduce the AIC values. The negative $\Delta$AIC increments are also relatively large compared to those in Table~\ref{tab: performance}. These improvements should not be exaggerated, however. The adjusted $R^2$ values from the OLS regressions still do not reach the $0.3$ threshold, for instance.

The second check relates to the classification of low-delay and high-delay groups (see Subsection~\ref{subsec: estimation}). The split according to median results in a roughly balanced metric for classification: $2,904$ of the CVEs observed attain a delay less than or equal to the median delay and $2,876$ a delay higher than the median delay of $15$ days. The accuracy of classifying these two groups is shown in Table~\ref{tab: classification performance}. The accuracy rates increase steadily up to the fourth model and decrease thereafter. These classification results agree with the regression results shown in Table~\ref{tab: performance}.

\begin{table}[th!b]
\centering
\caption{Classification Performance (median split)}
\label{tab: classification performance}
\begin{small}
\renewcommand*{\arraystretch}{1.2}
\begin{tabular}{lrrrrrr}
\hline
& \multicolumn{6}{c}{Model} \\
\cmidrule{2-7}
& $\Model_1$ & $\Model_2$ & $\Model_3$ & $\Model_4$ & $\Model_5$ & $\Model_6$ \\
\hline
Accuracy & 0.62 & 0.67 & 0.71 & 0.75 & 0.73 & 0.73 \\
\hline
\end{tabular}
\end{small}
\end{table}

\begin{table}[th!b]
\centering
\caption{OLS Performance in Annual Subsets (adj.~$R^2$)}
\label{tab: subset performance OLS}
\begin{small}
\renewcommand*{\arraystretch}{1.2}
\begin{tabular}{lrrrrrr}
\hline
& \multicolumn{6}{c}{Subset Model} \\
\cmidrule{2-7}
Year & $\AdjModel_1$ & $\AdjModel_2$ & $\AdjModel_3$ & $\AdjModel_4$ & $\AdjModel_5$ & $\AdjModel_6$ \\
\hline
2008 & 0.10 & 0.32 & 0.40 & 0.40 & 0.40 & 0.40 \\
2009 & 0.21 & 0.45 & 0.48 & 0.48 & 0.48 & 0.48 \\
2010 & 0.11 & 0.15 & 0.20 & 0.21 & 0.22 & 0.22 \\
2011 & 0.15 & 0.22 & 0.34 & 0.34 & 0.38 & 0.38 \\
2012 & 0.28 & 0.31 & 0.32 & 0.33 & 0.34 & 0.34 \\
2013 & 0.07 & 0.32 & 0.42 & 0.42 & 0.44 & 0.45 \\
2014 & 0.23 & 0.22 & 0.25 & 0.25 & 0.25 & 0.26 \\
2015 & 0.15 & 0.17 & 0.18 & 0.18 & 0.19 & 0.21 \\
2016 & 0.03 & 0.05 & 0.16 & 0.16 & 0.16 & 0.20 \\
\hline
~~$k$ & 13 & 17 & 23 & 26 & 29 & 39 \\
\hline
\end{tabular}
\end{small}
\end{table}

\begin{figure}[th!b]
\centering
\includegraphics[width=\linewidth, height=2.5cm]{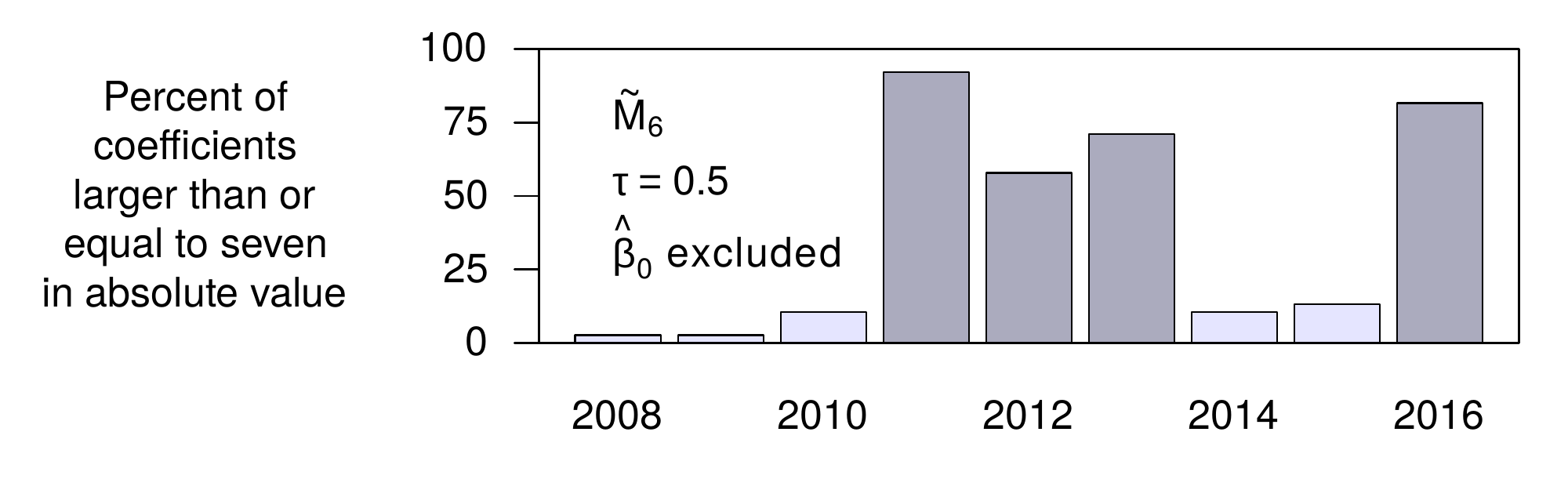}
\caption{Large Effects in Annual Subsets (QR, $\AdjModel_6$, $\tau = 0.5$)}
\label{fig: subset performance QR}
\end{figure}

\begin{figure}[th!b]
\centering
\includegraphics[width=\linewidth, height=8cm]{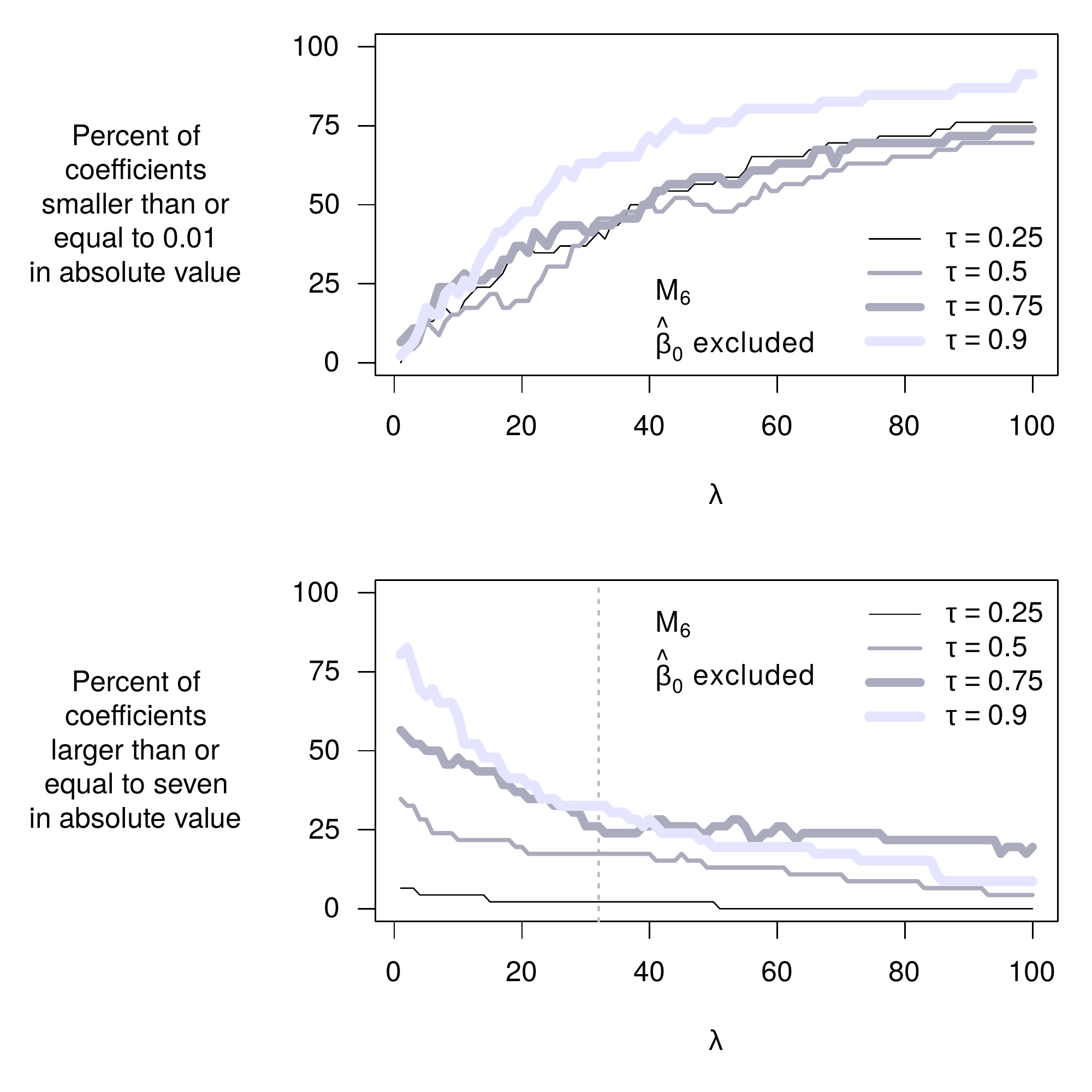}
\caption{QR Effects with LASSO ($\Model_6$)}
\label{fig: qr lasso}
\end{figure}

The third check is about the annual variation. In contrast to subsetting the explained metric, segmenting a model matrix into subsets according to the conditioning explanatory metrics is a sensible regression modeling approach~\cite{Koenker01}. Thus, to examine whether also the model performance varies annually, all six models are estimated with OLS in annual subsets, omitting the eight yearly dummy variables present in the original model specifications. For instance: if $\AdjModel_1$ denotes the first subset-model without the annual dummy variables, there are $k = 21 - 8 = 13$ parameters in the model and its regression coefficient vector. The results from these subset OLS regressions are summarized in Table~\ref{tab: subset performance OLS}. The table should be only read horizontally by comparing values in a given row across the columns; the different sample sizes in the annual subsets (see Fig.~\ref{fig: delays annual}) do not allow to make sound comparisons across the years. Given this interpretation guide, the results are clear. For the majority of years, there are no notable performance gains after the model $\AdjModel_3$. Another way to look at the annual variation is to check how well the large effect sizes of the annual dummy variables (see Fig.~\ref{fig: coefficients}) balance the effects of the other metrics. For this purpose, the median QR regression for the subset model $\AdjModel_6$ can be briefly examined. Even when keeping in mind that the coefficients are not directly comparable as some of the metrics have different scales, the summary shown in Fig.~\ref{fig: subset performance QR} roughly tells that the large effects are indeed pronounced during those years when the CVE coordination was delayed. Given that the majority of metrics show strong effects during these four years, the annual dummy variables seem to control relatively well the coefficient magnitudes reported in Fig.~\ref{fig: coefficients}.

The fourth and final check is about \eqref{eq: qr lasso}. The four hundred QR-LASSO regressions estimated are summarized in Fig.~\ref{fig: qr lasso} for $\Model_6$. \change{The upper plot indicates that} even with a modest regularization such as $\lambda = 20$, about a quarter of the coefficients are close to zero. Given that a week seems like a reasonably strong effect size with practical relevance, the lower plot indicates that the regularization seems to stabilize around $\lambda \leq 40$ for $\tau < 0.9$. For the $\tau = 0.9$ regressions, which generally gather the strongest effects (see Fig.~\ref{fig:  coefficients}), the applicable regularization seems to continue further. For the conditional median QR-LASSO regressions, only eight coefficients are larger than or equal to seven in absolute value at $\lambda = 30$. These are: the annual dummy variables for 2011, 2012, 2013, and 2016, the monthly February dummy variable, WEEKEND, SOCDEG, and NVDREFS. These belong to the longitudinal control metric group, the group of social network and communication metrics, and the infrastructure metric group. None of the CVSS and CWE metrics pass this subjective threshold.

\section{Discussion and Conclusions}\label{section: discussion}

The results presented allow to answer positively to each three research questions \ref{rq: social}, \ref{rq: infrastructure}, and \ref{rq: technical}. If the answers should be ordered in terms of importance of the corresponding statistical effects, the order would follow the numbering of the questions. The remainder of this paper summarizes the main findings, discusses the limitations, and points out a couple of prolific but challenging research paths for further work.

\subsection{Main Findings}

The main empirical findings can be summarized and theorized with the following five points.

\begin{enumerate}
\item{The first point is clear: even though nearly fifty explanatory metrics were considered, the statistical performance was relatively modest for explaining the CVE coordination delays. Less than one third of the total variation in the delays is explained by the OLS regression models estimated. The median quantile regression results largely agree. Analogous results were also obtained by classifying CVEs with a low-delay and high-delay split according to median. The adjusted coefficient of determination is not an ideal statistic to make comparisons (especially when OLS is not used), but it is still worth remarking that the bug tracking research frequently attains values around $0.5$ or even more \cite{Bhattacharya11, daCosta18}. The modest performance should not be interpreted as poor performance, however. Given the typical effect sizes seen in empirical software engineering research \cite{Kampenes08}, some of the regression coefficients and their standard errors indicate consistent, accurate, and large effects upon the CVE coordination delays.}
\item{The second point is likewise clear: most of the explanatory power comes from metrics used to proxy longitudinal variation. This result is not surprising. In contrast, it would be surprising if a software engineering study would not reveal longitudinal effects in a period covering almost a decade. The result is also familiar from comparable studies about vulnerability coordination~\cite{Ruohonen19ACI} and time series aspects of vulnerability archiving \cite{JohMalaiya17, Tang19}. If avoiding delays is important, the results can be also used to conjecture that ``do not request CVEs during weekends''. The longitudinal variation implies a further important takeaway message. Given that the efficiency of software engineering coordination is always dependent also on the \textit{volume} of items being coordinated, it seems that the \texttt{oss-security} case studied reflects the larger coordination issues that have gained publicity in recent years. By assumption, delays for CVE assignments are largely explained by the mere amount of these vulnerability identifiers requested. By implication, a good theoretical metric for prediction would simply be the amount of identifiers in an abstract CVE backlog.}
\item{The third point is about \textit{noise} that tends to lengthen the coordination delays. Both the social network (\ref{rq: social}) and the infrastructure (\ref{rq: infrastructure}) effects manifest themselves through different abstractions for such noise. In terms of the former, noise increases when there are ``too many cooks'' posting emails with high-entropy content. Both factors tend to increase also the CVE coordination delays. In the terms of the latter, noise increases with emails containing multiple hyperlinks to distinct tracking infrastructures within which the vulnerabilities have already been discussed or to which these have already been archived.}
\item{The fourth point relates to prerequisite constraints. When satisfied, these constraints tend to balance the positive effect of noise upon the delays by shortening the coordination delays to some extent. For instance, traces to bug tracking systems show a small but visible negative effect. The same applies regarding traces about the information sources regularly polled by the parties involved in the CVE tracking. For developers requesting CVE identifiers for their projects, the practical takeaway message is conveyed by an advice such as: ``write good bug reports to backup CVE requests''.}
\item{The fifth point relates to the technical characteristics of the vulnerabilities coordinated. The social network, communication, and infrastructure metrics (\ref{rq: social} and \ref{rq: infrastructure}) explain a few percentage points of the total variation in the delays. Given roughly comparable observations about the impact of social interactions on software quality~\cite{BettenburgHassan13}, these effects seem reasonable. Although the results are somewhat mixed, the technical characteristics (\ref{rq: technical}) do not generally explain the delays well. When backtracking to Fig.~\ref{fig: core}, the explanation may be simple: for participants who have assigned nearly thousand CVEs via \texttt{oss-security} alone, it may be irrelevant whether a request is about XSS or about buffer overflows. As there are still some signals about the statistical relevance of the CVSS and CWE metrics, another plausible theoretical explanation may be that it is mentally easier to handle vulnerabilities that fall explicitly into the scopes of some particular CWEs. Given the common problem of overloading a single vulnerability report with multiple distinct (security) issues \cite{Christey13}, the final practical takeaway message could be that: ``try to avoid ambiguities when requesting CVE identifiers''.}
\end{enumerate}

All in all, software vulnerability coordination can be concluded to exhibit typical characteristics of software engineering coordination in general. Dependencies, social relations, communication, and technical software elements are all present. These are also the factors that with varying degrees correlate with different coordination problems, including the CVE coordination delays observed.

\subsection{Threats to Validity}

Some limitations must be mentioned. According to a common taxonomy, the validity of the results reported are potentially exposed to threats to external validity, construct validity, and internal validity~\cite{Wohlin12}. The discussion that follows is structured according to this taxonomy.

\subsubsection{External Validity}

External validity relates to generalizability. Insofar as CVE assignments are considered, \texttt{oss-security} is a unique case that was limited to one particular way of assigning CVE identifiers during one particular historical period. Thus, neither the results nor the conclusions necessarily apply to other ways to assign CVEs, including the contemporary practices. While acknowledging this limitation, it is important to underline that analogous threats to external validity presumably continue to constrain also further research on vulnerability coordination.

On one hand, research in this domain is dependent on the openness (or lack thereof) of the internal tracking infrastructures used by MITRE and related parties. While keeping in mind the sensitivity of the information tracked, one question for practitioners to consider is whether data could be partially opened by using embargo periods akin to the grace periods used during vulnerability disclosure. The so-called distributed weakness filing project is a good step toward this direction~\cite{Openwall17a}. On the other hand, empirical research is more generally constrained by the lack of robust open data on the software engineering activities related to vulnerability coordination. The point applies also in the open source context. For instance, invitation-only coordination media \cite{Openwall17b} and other types of information hiding restrict the possibility of continuing the work \cite{Adams16} on security-related integration work done in open source projects. These constraints imply that further software engineering research in the domain is presumably as much about ``reverse engineering'' as it is about software and security engineering processes and coordination practices.

\subsubsection{Construct Validity}

The reverse engineering tenet is visible also in terms of construct validity. This type of validity relates to questions about how well the metrics used and the questions asked reflect the theoretical ideas and research goals. The most notable construct validity threat is directly related to the operationalization in~\eqref{eq: delays}. Because CVE requests were not explicitly modeled due to data limitations, even with the restriction in \eqref{eq: delays restriction}, it is impossible to say whether a CVE that appeared on the list was already disseminated to MITRE via other channels. While the delays observed should still provide a reasonable approximation, it is perhaps more important to note that the metric used provides only a limited viewpoint on the coordination. Unfortunately, the lack of open data makes it difficult to study more nuanced coordination aspects such as task allocation and work parallelism related to CVE coordination. Bayesian methods and expert opinion \cite{Johnson17, Johnston18} may help at resolving this constraint and related limitations.

Also some of the explanatory metrics are exposed to modest construct validity threats. Without attempting to participate in the current debates \cite{Allodi14, Johnson17, YonisMalaiya16}, it is reasonable to assert that the CVSS and CWE data from NVD is robust enough for the purposes of this paper. Construct validity is a bigger issue for the few custom metrics derived. Most of the social network and communication metrics have been successfully used previously~\cite{BettenburgHassan13, Bird11}. The operationalization was also deliberately restricted to metrics that are easy to compute and interpret. Therefore, the approximations used for the infrastructure metrics are more noteworthy. In particular, the regular expressions in Table~\ref{tab: domain regex} are inadequate for explicitly linking different tracking infrastructures together. While a subset of the CVEs discussed on \texttt{oss-security} could be explicitly linked to bug tracking and related systems via repository mining techniques \cite{DashevskyiMassacci19, Massacci13, Romo14}, the historical period studied largely prevents such linking because many of the trackers are nonexistent today (see Subsection~\ref{subsec: infrastructure metrics}). In this sense, the approximations are a necessary evil for this paper.

\subsubsection{Internal Validity}

Internal validity relates to questions about computational and statistical biases particularly when causal relations are postulated. Three internal validity threats are worth pointing out. First, even when causal inference is not attempted, the data availability issues translate to potential problems in terms of omitted variables and confounding factors. Second, not all options were examined for thoroughly assessing the reasons behind the modest performance. For instance, clustering could be used in conjunction with regression analysis~\cite{Bettenburg14, Ruohonen16AICCSA}. Another option might be to examine different CWE-based ontologies~\cite{Tsipenyuk05, WuGandhi10}. The third and final threat to internal validity relates to the statistical issues reported.

As is typical to software engineering datasets~\cite{Fenton99, Kitchenham17}, non-normality, heteroskedasticity, and multicollinearity were all present with varying degrees of severity. Further work is required to examine these issues with respect to generally more robust but still not invulnerable techniques such as quantile regression and LASSO. Quantile regression addresses the distributional assumption, while the heteroskedasticity patterns observed (see Fig.~\ref{fig: qr residuals}) would be easy to explicitly adjust for in further work. The multicollinearity issue is more interesting. Given that the CVSS (v.~2) metrics are alone challenging to adequately handle in empirical research due to the various plausible combinations between the variables, further work is also required on variable selection in the software engineering context. The potential solutions should be systematic and algorithmic to ensure consistency of theoretical arguments---otherwise the so-called researcher bias is inevitable for regression analysis with interacting variables. By using combinations of CVSS metrics, combinations of CWE metrics, combinations of CVSS and CWE metrics, and other combinations, it would be possible to assert almost an unlimited amount of theoretical propositions.

Finally, the real issue is more subtle than what might be remedied by merely changing a statistical modeling approach. There have been historical delays also for CVSS assignments \cite{Ruohonen19ACI}, there are delays between reporting bugs in tracking systems, fixing these in version control systems, and integrating the fixes to releases \cite{daCosta18, RodriguezPerez17}, and so forth. All such delays may affect also CVE coordination. While statistical modeling is one thing, analytical understanding of the complexities involved is another.

\subsection{Moving Forward}

The limitations discussed prompt a couple of points about potential means for moving forward. First, it is sensible to recommend a closer marriage between the software vulnerability and bug tracking research domains. The intersection between the domains is large but seldom explicitly articulated. Such a marriage might also remedy the \change{somewhat} modest statistical performance reported. \change{In this regard, a sensible conclusion is that the} information used was too limited for capturing the truly relevant elements affecting the particular type of coordination observed. When looking at the bug tracking research domain, some of the frequently incorporated dimensions (such as the reputation of bug reporters and the text mining of bug reports) seem prolific for pursuing further also in the context of software vulnerabilities. These and related dimensions are relevant for studying also more contemporary questions such as those related to bug bounties.

Second, a marriage between two domains alone seems insufficient for making more practical advances. Mining of software repositories frequently provides valuable insights about software engineering coordination through case studies. Yet, as was briefly demonstrated, the practical problem is that there are hundreds (if not thousands) of relevant repositories to mine. For advances with practical relevance, affairs are required also with other computer science domains such as information retrieval and web crawling. If one is to believe the current ``mega trends", maybe some day, perhaps in a galaxy far away, also the coordination of abstract identifiers could be done by robots.

\balance
\bibliographystyle{apalike}

\begin{thebibliography}{}

\bibitem[Ablon and Bogart, 2017]{Ablon17}
Ablon, L. and Bogart, A. (2017).
\newblock {Z}ero {D}ays, {T}housands of {N}ights: {T}he {L}ife and {T}imes of
  {Z}ero-{D}ay {V}ulnerabilities and {T}heir {E}xploits.
\newblock {RAND} {C}orporation, {S}anta {M}onica. {A}vailable online in
  September 2017:
  \url{https://www.rand.org/content/dam/rand/pubs/research_reports/RR1700/RR1751/RAND_RR1751.pdf}.

\bibitem[Adams et~al., 2016]{Adams16}
Adams, B., Kavanagh, R., Hassan, A.~E., and German, D.~M. (2016).
\newblock {A}n {E}mpirical {S}tudy of {I}ntegration {A}ctivities in
  {D}istributions of {O}pen {S}ource {S}oftware.
\newblock {\em Empirical Software Engineering}, 21(3):960--1001.

\bibitem[Alhamzawi et~al., 2012]{Alhamzawi12}
Alhamzawi, R., Yu, K., and Benoit, D.~F. (2012).
\newblock {B}ayesian {A}daptive {L}asso {Q}uantile {R}egression.
\newblock {\em Statistical Modelling}, 12(3):\text{279--297}.

\bibitem[Allodi, 2017]{Allodi17c}
Allodi, L. (2017).
\newblock {E}conomic {F}actors of {V}ulnerability {T}rade and {E}xploitation:
  {E}mpirical {E}vidence from a {P}rominent {R}ussian {C}ybercrime {M}arket.
\newblock In {\em Proceedings of the ACM Conference on Computer and
  Communications Security (CCS 2017)}, pages 1483--1499, Dallas. ACM.

\bibitem[Allodi et~al., 2017]{Allodi17d}
Allodi, L., Biagioni, S., Crispo, B., Labunets, K., Massacci, F., and Santos,
  W. (2017).
\newblock {E}stimating the {A}ssessment {D}ifficulty of {CVSS} {E}nvironmental
  {M}etrics: {A}n {E}xperiment.
\newblock In Dang, T.~K., Wagner, R., K{\"u}ng, J., Thoai, N., Takizawa, M.,
  and Neufhold, E.~J., editors, {\em Proceedings of the International
  Conference on Future Data and Security Engineering (FDSE 2017), Lecture Notes
  in Computer Science (Volume 10646)}, pages 23--39. Springer.

\bibitem[Allodi and Massacci, 2014]{Allodi14}
Allodi, L. and Massacci, F. (2014).
\newblock {C}omparing {V}ulnerability {S}everity and {E}xploits {U}sing
  {C}ase-{C}ontrol {S}tudies.
\newblock {\em ACM Transactions on Information and System Security},
  17(1):1:1--1:20.

\bibitem[Arora et~al., 2010]{Arora10}
Arora, A., Forman, C., Nandkumar, A., and Telang, R. (2010).
\newblock {C}ompetition and {P}atching of {S}ecurity {V}ulnerabilities: {A}n
  {E}mpirical {A}nalysis.
\newblock {\em Information Economics and Policy}, 22(2):164--177.

\bibitem[Bettenburg and Hassan, 2013]{BettenburgHassan13}
Bettenburg, N. and Hassan, A.~E. (2013).
\newblock {S}tudying the {I}mpact of {S}ocial {I}nteractions on {S}oftware
  {Q}uality.
\newblock {\em Empirical Software Engineering}, 18(2):375--431.

\bibitem[Bettenburg et~al., 2014]{Bettenburg14}
Bettenburg, N., Nagappan, M., and Hassan, A.~E. (2014).
\newblock {T}owards {I}mproving {S}tatistical {M}odeling of {S}oftware
  {E}ngineering {D}ata: {T}hink {L}ocally, {A}ct {G}lobally!
\newblock {\em Empirical Software Engineering}, 20(2):294--335.

\bibitem[Bhattacharya and Neamtiu, 2011]{Bhattacharya11}
Bhattacharya, P. and Neamtiu, I. (2011).
\newblock {B}ug-{F}ix {T}ime {P}rediction {M}odels: {C}an {W}e {D}o {B}etter?
\newblock In {\em Proceedings of the 8th Working Conference on Mining Software
  Repositories (MSR 2011)}, pages 207--210, Waikiki. ACM.

\bibitem[Bilge and Dumitras, 2012]{Bilge12}
Bilge, L. and Dumitras, T. (2012).
\newblock {B}efore {W}e {K}new {I}t: {A}n {E}mpirical {S}tudy of {Z}ero-{D}ay
  {A}ttacks in the {R}eal {W}orld.
\newblock In {\em Proceedings of the 2012 ACM Conference on Computer and
  Communications Security (CCS 2012)}, pages 833--844, Raleigh. ACM.

\bibitem[Bird, 2011]{Bird11}
Bird, C. (2011).
\newblock {S}ociotechnical {C}oordination and {C}ollaboration in {O}pen
  {S}ource {S}oftware.
\newblock In {\em Proceedings of the 2011 27th IEEE International Conference on
  Software Maintenance (ICSM 2011)}, pages 568--573, Williamsburg. IEEE.

\bibitem[Bird et~al., 2006]{Bird06}
Bird, C., Gourley, A., Devanbu, P., Gertz, M., and Swaminathan, A. (2006).
\newblock {M}ining {E}mail {S}ocial {N}etworks.
\newblock In {\em Proceedings of the 2006 International Workshop on Mining
  Software Repositories (MSR 2006)}, pages 137--143, Shanghai. ACM.

\bibitem[Blincoe et~al., 2015]{Blincoe15}
Blincoe, K., Valetto, G., and Damian, D. (2015).
\newblock {F}acilitating {C}oordination between {S}oftware {D}evelopers: {A}
  {S}tudy and {T}echniques for {T}imely and {E}fficient {R}ecommendations.
\newblock {\em IEEE Transactions on Software Engineering}, 41(10):969--985.

\bibitem[Bozorgi et~al., 2010]{Bozorgi10}
Bozorgi, M., Saul, L.~K., Savage, S., and Voelker, G.~M. (2010).
\newblock {B}eyond {H}euristics: {L}earning to {C}lassify {V}ulnerabilities and
  {P}redict {E}xploits.
\newblock In {\em Proceedings of the 16th ACM SIGKDD International Conference
  on Knowledge Discovery and Data Mining (KDD 2010)}, pages 105--114, London.
  ACM.

\bibitem[Cetin et~al., 2017]{CetinVanEeeten17}
Cetin, O., {Ga\~{n}\'{a}n}, C., Korczy\'nski, M., and {van Eeten}, M. (2017).
\newblock {M}ake {N}otifications {G}reat {A}gain: {L}earning {H}ow to {N}otify
  in the {A}ge of {L}arge-{S}cale {V}ulnerability {S}canning.
\newblock In {\em Proceedings of the 16th Workshop on the Economics of
  Information Security (WEIS 2017)}, San Diego.
\newblock {A}vailable online in January 2018:
  \url{http://weis2017.econinfosec.org/wp-content/uploads/sites/3/2017/05/WEIS_2017_paper_17.pdf}.

\bibitem[Chia-Yen~Lee, 2017]{LeeChen17}
Chia-Yen~Lee, B.-S.~C. (2017).
\newblock {M}utually-{E}xclusive-and-{C}ollectively-{E}xhaustive {F}eature
  {S}election {S}cheme.
\newblock {\em Applied Soft Computing}, 68:961--971.

\bibitem[Christen, 2012]{Christen12}
Christen, P. (2012).
\newblock {\em {D}ata {M}atching: {C}oncepts and {T}echniques for {R}ecord
  {L}inkage, {E}ntity {R}esolution, and {D}uplicate {D}etection}.
\newblock Springer, Berlin.

\bibitem[Christey and Martin, 2013]{Christey13}
Christey, S. and Martin, B. (2013).
\newblock {B}uying {I}nto the {B}ias: {W}hy {V}ulnerability {S}tatistics
  {S}uck.
\newblock In {\em Presentation at Black Hat 2013}, Las Vegas.
\newblock available online in January 2017:
  \url{https://media.blackhat.com/us-13/US-13-Martin-Buying-Into-The-Bias-Why-Vulnerability\\-Statistics-Suck-Slides.pdf}.

\bibitem[Christey and Wysopal, 2002]{Christey02}
Christey, S. and Wysopal, C. (2002).
\newblock {R}esponsible {V}ulnerability {D}isclosure {P}rocess.
\newblock {Internet Engineering Task Force (IETF)}, {INTERNET-DRAFT}, available
  online in December 2017:
  \url{https://tools.ietf.org/html/draft-christey-wysopal-vuln-disclosure-00}.

\bibitem[Conaldi and Lomi, 2013]{Conaldi13}
Conaldi, G. and Lomi, A. (2013).
\newblock {T}he {D}ual {N}etwork {S}tructure of {O}rganizational {P}roblem
  {S}olving: {A} {C}ase {S}tudy on {O}pen {S}ource {S}oftware {D}evelopment.
\newblock {\em Social Networks}, 35(2):237--250.

\bibitem[Conaldi et~al., 2012]{Conaldi12}
Conaldi, G., Lomi, A., and Tonellato, M. (2012).
\newblock {D}ynamic {M}odels of {A}ffiliation and the {N}etwork {S}tructure of
  {P}roblem {S}olving in {O}pen {S}ource {S}oftware {P}rojects.
\newblock {\em Organizational Research Methods}, 15(3):385--412.

\bibitem[Conway, 1968]{Conway68}
Conway, M.~E. (1968).
\newblock {H}ow {D}o {C}ommittees {I}nvent?
\newblock {\em Datamation}, 14(5):28--31.

\bibitem[Crowston and Howison, 2006]{Crowston06}
Crowston, K. and Howison, J. (2006).
\newblock {H}ierarchy and {C}entralization in {F}ree and {O}pen {S}ource
  {S}oftware {T}eam {C}ommunications.
\newblock {\em Knowledge, Technology \& Policy}, 18(4):65--85.

\bibitem[Crowston and Shamshurin, 2017]{Crowston17}
Crowston, K. and Shamshurin, I. (2017).
\newblock {C}ore-{P}eriphery {C}ommunication and the {S}uccess of
  {F}ree/{L}ibre {O}pen {S}ource {S}oftware {P}rojects.
\newblock {\em Journal of Internet Services and Applications},
  8(10):\text{1--11}.

\bibitem[{da Costa} et~al., 2018]{daCosta18}
{da Costa}, D.~A., McIntosh, S., Kulesza, U., Hassan, A.~E., and Abebe, S.~L.
  (2018).
\newblock {A}n {E}mpirical {S}tudy of the {I}ntegration {T}ime of {F}ixed
  {I}ssues.
\newblock {\em Empirical Software Engineering}, 23(1):334--383.

\bibitem[Dashevskyi et~al., 2019]{DashevskyiMassacci19}
Dashevskyi, S., Brucker, A.~D., and Massacci, F. (2019).
\newblock {A} {S}creening {T}est for {D}isclosed {V}ulnerabilities in {FOSS}
  {C}omponents.
\newblock {\em IEEE Transactions on Software Engineering}, 45(10):945--966.

\bibitem[Dehghani et~al., 2012]{Dehghani12}
Dehghani, M., Asadpour, M., and Shakery, A. (2012).
\newblock {A}n {E}volutionary-{B}ased {M}ethod for {R}econstructing
  {C}onversation {T}hreads in {E}mail {C}orpora.
\newblock In {\em Proceedings of the IEEE/ACM International Conference on
  Advances in Social Networks Analysis and Mining (ASONAM 2012)}, pages
  \text{1132--1137}, Istanbul. IEEE.

\bibitem[Eyolfson et~al., 2011]{Eyolfson11}
Eyolfson, J., Tan, L., and Lam, P. (2011).
\newblock {D}o {T}ime of {D}ay and {D}eveloper {E}xperience {A}ffect {C}ommit
  {B}ugginess?
\newblock In {\em Proceedings of the 8th Working Conference on Mining Software
  Repositories (MSR 2011)}, pages 153--162, Honolulu. ACM.

\bibitem[Fenton and Neil, 1999]{Fenton99}
Fenton, N.~E. and Neil, M. (1999).
\newblock {A} {C}ritique of {S}oftware {D}efect {P}rediction {M}odels.
\newblock {\em IEEE Transactions on Software Engineering}, 25(5):675--689.

\bibitem[FIRST, 2007]{FIRST07}
FIRST (2007).
\newblock {A} {C}omplete {G}uide to the {C}ommon {V}ulnerability {S}coring
  {S}ystem {V}ersion {2.0}, {FIRST.ORG}.
\newblock Available online in June 2015:
  \url{https://www.first.org/cvss/cvss-v2-guide.pdf}.

\bibitem[Giacalone et~al., 2017]{Giacalone17}
Giacalone, M., Panarello, D., and Mattera, R. (2017).
\newblock {M}ulticollinearity in {R}egression: {A}n {E}fficiency {C}omparison
  {B}etween \text{L$_p$-Norm} and {L}east {S}quares {E}stimators.
\newblock {\em Quality \& Quantity}, 52(4):1831--1859.

\bibitem[{Goseva-Popstojanova} and Perhinschi, 2015]{GosevaPopstojanova15}
{Goseva-Popstojanova}, K. and Perhinschi, A. (2015).
\newblock {O}n the {C}apability of {S}tatic {C}ode {A}nalysis to {D}etect
  {S}ecurity {V}ulnerabilities.
\newblock {\em Information and Software Technology}, 68:17--33.

\bibitem[Guzzi et~al., 2013]{Guzzi13}
Guzzi, A., Bacchelli, A., Lanza, M., Pinzger, M., and {van Deursen}, A. (2013).
\newblock {C}ommunication in {O}pen {S}ource {S}oftware {D}evelopment {M}ailing
  {L}ist.
\newblock In {\em Proceedings of the 10th Working Conference on Mining Software
  Repositories (MSR 2013)}, pages \text{277--286}, San Francisco. IEEE.

\bibitem[Habayeb et~al., 2018]{Habayeb18}
Habayeb, M., Murtaza, S.~S., Miranskyy, A., and Bener, A.~B. (2018).
\newblock {O}n the {U}se of {H}idden {M}arkov {M}odel to {P}redict the {T}ime
  to {F}ix {B}ugs.
\newblock {\em IEEE Transactions on Software Engineering}, 44(12):1224--1244.

\bibitem[Howison and Crowston, 2014]{HowisonCrowston14}
Howison, J. and Crowston, K. (2014).
\newblock {C}ollaboration {T}hrough {O}pen {S}uperposition: {A} {T}heory of the
  {O}pen {S}ource {W}ay.
\newblock {\em MIS Quarterly}, 38(1):\text{29--50}.

\bibitem[Joh and Malaiya, 2017]{JohMalaiya17}
Joh, H. and Malaiya, Y.~K. (2017).
\newblock {P}eriodicity in {S}oftware {V}ulnerability {D}iscovery, {P}atching
  and {E}xploitation.
\newblock {\em International Journal of Information Security}, 16(6):673--690.

\bibitem[Johnson and Ekstedt, 2016]{Johnson16a}
Johnson, P. and Ekstedt, M. (2016).
\newblock {T}he {T}arpit -- {A} {G}eneral {T}heory of {S}oftware {E}ngineering.
\newblock {\em Information and Software Technology}, 70:181--203.

\bibitem[Johnson et~al., 2016]{Johnson16b}
Johnson, P., Gorton, D., Langerstr\"om, R., and Ekstedt, M. (2016).
\newblock {T}ime {B}etween {V}ulnerability {D}isclosures: {A} {M}easure of
  {S}oftware {P}roduct {V}ulnerability.
\newblock {\em Computers \& Security}, 62:\text{278--295}.

\bibitem[Johnson et~al., 2017]{Johnson17}
Johnson, P., Lagerstr\"om, R., Ekstedt, M., and Franke, U. (2017).
\newblock {C}an the {C}ommon {V}ulnerability {S}coring {S}ystem be {T}rusted?
  {A} {B}ayesian {A}nalysis.
\newblock {\em IEEE Transactions on Dependable and Secure Computing},
  15(6):1002--1015.

\bibitem[Johnston et~al., 2018]{Johnston18}
Johnston, R., Sarkani, S., Mazzuchi, T., Holzer, T., and Eveleigh, T. (2018).
\newblock {M}ultivariate {M}odels {U}sing {MCMCB}ayes for {W}eb-{B}rowser
  {V}ulnerability {D}iscovery.
\newblock {\em Reliability Engineering \& System Safety}, 176:52--61.

\bibitem[Kampenes et~al., 2008]{Kampenes08}
Kampenes, V.~B., Dyb{\aa}, T., Hannay, J.~E., and Sj{\o}berg, D.~I. (2008).
\newblock {A} {S}ystematic {R}eview of {E}ffect {S}ize in {S}oftware
  {E}ngineering {E}xperiments.
\newblock {\em Information and Software Technology}, 49(11--12):1073--1086.

\bibitem[Karlsson et~al., 2019]{Karlsson19}
Karlsson, P.~S., Behrenz, L., and Shukur, G. (2019).
\newblock {P}erformance of {M}odel {S}election {C}riteria {W}hen {V}ariables
  are {I}ll {C}onditioned.
\newblock {\em Computational Economics}, 54:\text{77--98}.

\bibitem[Kitchenham et~al., 2017]{Kitchenham17}
Kitchenham, B., Madeyski, L., Budgen, D., Keung, J., Brereton, P., Charters,
  S., Gibbs, S., and Pohthong, A. (2017).
\newblock {R}obust {S}tatistical {M}ethods for {E}mpirical {S}oftware
  {E}ngineering.
\newblock {\em Empirical Software Engineering}, 22(2):579--630.

\bibitem[Koenker and Bassett, 1982]{Koenker82}
Koenker, R. and Bassett, G. (1982).
\newblock {R}obust {T}ests for {H}eteroscedasticity {B}ased on {R}egression
  {Q}uantiles.
\newblock {\em Econometrica}, 50(1):43--61.

\bibitem[Koenker and Hallock, 2001]{Koenker01}
Koenker, R. and Hallock, K.~F. (2001).
\newblock {Q}uantile {R}egression.
\newblock {\em Journal of Economic Perspectives}, 15(4):143--156.

\bibitem[Koenker and Machado, 1999]{Koenker99}
Koenker, R. and Machado, J. A.~F. (1999).
\newblock {G}oodness of {F}it and {R}elated {I}nference {P}rocesses for
  {Q}uantile {R}egression.
\newblock {\em Journal of the American Statistical Association},
  94(448):1296--1310.

\bibitem[Koenker et~al., 2018]{quantreg18}
Koenker, R., Portnoy, S., Ng, P.~T., Zeileis, A., Grosjean, P., and Ripley,
  B.~D. (2018).
\newblock {quantreg}: {Q}uantile {R}egression.
\newblock {R} package version 5.3.5, available online in April 2018:
  \url{https://cran.r-project.org/web/packages/quantreg/index.html}.

\bibitem[Kuhn, 2008]{caret}
Kuhn, M. (2008).
\newblock {B}uilding {P}redictive {M}odels in {R} {U}sing the \texttt{caret}
  {P}ackage.
\newblock {\em Journal of Statistical Software}, 28(5):1--26.

\bibitem[Kuk, 2006]{Kuk06}
Kuk, G. (2006).
\newblock {S}trategic {I}nteraction and {K}nowledge {S}haring in the {KDE}
  {D}eveloper {M}ailing {L}ist.
\newblock {\em Management Science}, 52(7):\text{1031--1042}.

\bibitem[Kula et~al., 2010]{Kula10}
Kula, R.~G., Fushida, K., Kawaguchi, S., and Iida, H. (2010).
\newblock {A}nalysis of {B}ug {F}ixing {P}rocesses {U}sing {P}rogram {S}licing
  {M}etrics.
\newblock In Babar, M.~A., Vierimaa, M., and Oivo, M., editors, {\em
  Proceedings of the International Conference on Product Focused Software
  Process Improvement (PROFES 2010), Lecture Notes in Computer Science (Volume
  6156)}, pages 32--46, Limerick. Springer.

\bibitem[Laszka et~al., 2016]{Laszka16}
Laszka, A., Zhao, M., and Grossklags, J. (2016).
\newblock {B}anishing {M}isaligned {I}ncentives for {V}alidating {R}eports in
  {B}ug-{B}ounty {P}latforms.
\newblock In Askoxylakis, I., Ioannidis, S., Katsikas, S., and Meadows, C.,
  editors, {\em Proceedings of the European Symposium on Research in Computer
  Security (ESORICS 2016), Lecture Notes in Computer Science (Volume 9879)},
  pages 161--178, Heraklion. Springer.

\bibitem[Lee et~al., 2013]{Lee13}
Lee, G., Espinosa, J.~A., and DeLone, W.~H. (2013).
\newblock {T}ask {E}nvironment {C}omplexity, {G}lobal {T}eam {D}ispersion,
  {P}rocess {C}apabilities, and {C}oordination in {S}oftware {D}evelopment.
\newblock {\em IEEE Transactions on Software Engineering}, 39(12):1753--1771.

\bibitem[Levenshtein, 1966]{Levenshtein66}
Levenshtein, V.~I. (1966).
\newblock {B}inary {C}odes {C}apable of {C}orrecting {D}eletions, {I}nsertions,
  and {R}eversals.
\newblock {\em Soviet Physics-Doklady}, 10(8):707--710.

\bibitem[Leyden, 2017]{Leyden17}
Leyden, J. (2017).
\newblock {M}ost {V}ulnerabilities {F}irst {B}labbed {A}bout {O}nline or on the
  {D}ark {W}eb: {O}fficial {B}ug {N}otice? {S}ure, but not {B}efore {I} {G}et
  {C}red and {LOLs}.
\newblock {T}he {R}egister. Available online in December 2017:
  \url{http://www.theregister.co.uk/2017/06/08/vuln_disclosure_lag/}.

\bibitem[Liaw and Wiener, 2002]{randomForest}
Liaw, A. and Wiener, M. (2002).
\newblock {C}lassification and {R}egression by {randomForest}.
\newblock {\em R News}, 2(3):18--22.

\bibitem[Licorish and MacDonell, 2014]{Licorish14}
Licorish, S.~A. and MacDonell, S.~G. (2014).
\newblock {U}nderstanding the {A}ttitudes, {K}nowledge {S}haring {B}ehaviors
  and {T}ask {P}erformance of {C}ore {D}evelopers: {A} {L}ongitudinal {S}tudy.
\newblock {\em Information and Software Technology}, 56(12):1578--1596.

\bibitem[{Linares-V\'asquez} et~al., 2017]{LinaresVasquez17}
{Linares-V\'asquez}, M., Bavota, G., and {Escobar-Vel\'asquez}, C. (2017).
\newblock {A}n {E}mpirical {S}tudy on {A}ndroid-{R}elated {V}ulnerabilities.
\newblock In {\em IEEE/ACM 14th International Conference on Mining Software
  Repositories (MSR 2017)}, pages 1--13, Buenos Aires. IEEE.

\bibitem[Lubarski and Morzy, 2012]{Lubarski12}
Lubarski, P. and Morzy, M. (2012).
\newblock {M}easuring the {I}mportance of {U}sers in a {S}ocial {N}etwork
  {B}ased on {E}mail {C}ommunication {P}atterns.
\newblock In {\em Proceedings of the IEEE/ACM International Conference on
  Advances in Social Network Analysis and Mining (ASONAM 2012)}, pages
  \text{86--90}, Istanbul. IEEE.

\bibitem[L\"utkepohl, 2007]{Lutkepohl07}
L\"utkepohl, H. (2007).
\newblock {G}eneral-to-{S}pecific or {S}pecific-to-{G}eneral {M}odelling? {A}n
  {O}pinion on {C}urrent {E}conometric {T}erminology.
\newblock {\em Journal of Econometrics}, 136(1):319--324.

\bibitem[MacKinnon and White, 1985]{MacKinnonWhite85}
MacKinnon, J.~G. and White, H. (1985).
\newblock {S}ome {H}eteroskedasticity-{C}onsistent {C}ovariance {M}atrix
  {E}stimators with {I}mproved {F}inite {S}ample {P}roperties.
\newblock {\em Journal of Econometrics}, 29(3):305--325.

\bibitem[Malone and Crowston, 1994]{MaloneCrowston94}
Malone, T.~W. and Crowston, K. (1994).
\newblock {T}he {I}nterdisciplinary {S}tudy of {C}oordination.
\newblock {\em ACM Computing Surveys}, 26(1):87--119.

\bibitem[McChesney, 1997]{McChesney97}
McChesney, I.~R. (1997).
\newblock {E}ffective {C}oordination in the {S}oftware {P}rocess --
  {H}istorical {P}erspectives and {F}uture {D}irections.
\newblock {\em Software Quality Journal}, 6(3):235--246.

\bibitem[McQueen et~al., 2009]{McQueen09}
McQueen, M.~A., McQueen, T.~A., Boyer, W.~F., and Chaffin, M.~R. (2009).
\newblock {E}mpirical {E}stimates and {O}bservations of 0{D}ay
  {V}ulnerabilities.
\newblock In {\em Proceedings of the 42nd Hawaii International Conference on
  System Sciences (HICSS 2009)}, pages 1--12, Honolulu. IEEE.

\bibitem[Meneely and Williams, 2010]{Meneely10}
Meneely, A. and Williams, L. (2010).
\newblock {S}trengthening the {E}mpirical {A}nalysis of the {R}elationship
  {B}etween {L}inus' {L}aw and {S}oftware {S}ecurity.
\newblock In {\em Proceedings of the 2010 ACM-IEEE International Symposium on
  Empirical Software Engineering and Measurement (ESEM 2010)}, pages 9:1--9:10,
  Bolzano-Bozen. ACM.

\bibitem[{MITRE}, 2015a]{MITRE15a}
{MITRE} (2015a).
\newblock {CVE-ID} {S}yntax {C}hange.
\newblock Available online in December 2017:
  \url{https://cve.mitre.org/cve/identifiers/syntaxchange.html}.

\bibitem[{MITRE}, 2015b]{MITRE15b}
{MITRE} (2015b).
\newblock {F}requently {A}sked {Q}uestions.
\newblock Available online in December 2017:
  \url{https://cve.mitre.org/about/faqs.html}.

\bibitem[{MITRE}, 2015c]{MITRE15c}
{MITRE} (2015c).
\newblock {P}lease welcome {K}urt {S}eifried to the {CVE} {E}ditorial {B}oard.
\newblock Appeared originally in \textit{cve-editorial-board-list}. Available
  online in September 2016:
  \url{https://cve.mitre.org/data/board/archives/2015-11/msg00002.html}.

\bibitem[{MITRE}, 2018a]{MITRE18a}
{MITRE} (2018a).
\newblock {C}ommon {W}eaknesses {E}numeration.
\newblock Available online in January 2018: \url{http://cwe.mitre.org/}.

\bibitem[{MITRE}, 2018b]{MITRE18b}
{MITRE} (2018b).
\newblock {CWE VIEW}: {W}eaknesses {O}riginally {U}sed by {NVD} from 2008 to
  2016.
\newblock Available online in January 2018:
  \url{http://cwe.mitre.org/data/definitions/635.html}.

\bibitem[Ngamkajornwiwat et~al., 2008]{Ngamkajornwiwat08}
Ngamkajornwiwat, K., Zhang, D., Koru, A.~G., Zhou, L., and Nolker, R. (2008).
\newblock {A}n {E}xploratory {S}tudy on the {E}volution of {OSS} {D}eveloper
  {C}ommunities.
\newblock In {\em Proceedings of the 41st Annual Hawaii International
  Conference on System Sciences (HICSS 2008)}, pages 305--315, Waikoloa. IEEE.

\bibitem[Nguyen and Massacci, 2013]{Massacci13}
Nguyen, V.~H. and Massacci, F. (2013).
\newblock {T}he ({U}n){R}eliability of {NVD} {V}ulnerability {V}ersions {D}ata:
  {A}n {E}mpirical {E}xperiment on {G}oogle {C}hrome {V}ulnerabilities.
\newblock In {\em {P}roceedings of the 8th ACM SIGSAC Symposium on Information,
  Computer and Communications Security (ASIACCS 2013)}, pages 493--498. ACM.

\bibitem[Nia et~al., 2010]{Nia10}
Nia, R., Bird, C., Devanbu, P., and Filkov, V. (2010).
\newblock {V}alidity of {N}etwork {A}nalyses in {O}pen {S}ource {P}rojects.
\newblock In {\em Proceedings of the 7th IEEE Working Conference on Mining
  Software Repositories (MSR 2010)}, pages 201--209, Cape Town. IEEE.

\bibitem[{NIST}, 2017]{NVD17a}
{NIST} (2017).
\newblock {NVD} {D}ata {F}eed and {P}roduct {I}ntegration.
\newblock {N}ational Institute of Standards and Technology~(NIST), Annually
  Archived CVE Vulnerability Feeds: Security Related Software Flaws, NVD/CVE
  XML Feed with CVSS and CPE Mappings (Version 2.0). Retrieved in 23 September
  2017 from: \url{https://nvd.nist.gov/download.cfm}.

\bibitem[{NIST}, 2018]{NVD18a}
{NIST} (2018).
\newblock {C}ommon {V}ulnerability {S}coring {S}ystem {C}alculator: {V}ersion 2
  -- \text{CVE-2017-5754}.
\newblock {N}ational Institute of Standards and Technology~(NIST), Available
  online in January 2018:
  \url{https://nvd.nist.gov/vuln-metrics/cvss/v2-calculator?name=CVE-2017-5754&vector=(AV:L/AC:M/Au:N/C:C/I:N/A:N)}.

\bibitem[{Openwall}, 2008]{Openwall08b}
{Openwall} (2008).
\newblock {CVE} request: mantisbt \textless 1.1.4: {RCE}.
\newblock Available online in January 2018:
  \url{http://openwall.com/lists/oss-security/2008/10/19/1}.

\bibitem[{Openwall}, 2016a]{Openwall16a}
{Openwall} (2016a).
\newblock {A}rchive of \textit{oss-security} {M}ailing {L}ist.
\newblock Available online in September 2016:
  \url{http://www.openwall.com/lists/oss-security/}.

\bibitem[{Openwall}, 2016b]{Openwall16c}
{Openwall} (2016b).
\newblock {F}wd: {CVE} request - samsumg android phone {SVE-2016-6244}
  {P}ossible {P}rivilege {E}scalation in telecom.
\newblock Available online in September 2016:
  \url{http://www.openwall.com/lists/oss-security/2016/08/05/1}.

\bibitem[{Openwall}, 2017a]{Openwall17a}
{Openwall} (2017a).
\newblock {MITRE is adding data intake to its CVE ID process}.
\newblock Available online in December 2017:
  \url{http://www.openwall.com/lists/oss-security/2017/02/09/7}.

\bibitem[{Openwall}, 2017b]{Openwall17b}
{Openwall} (2017b).
\newblock {Re: linux-distros subscription}.
\newblock Available online in December 2017:
  \url{http://www.openwall.com/lists/oss-security/2017/01/15/1}.

\bibitem[Paasivaara and Lassenius, 2003]{Paasivaara03}
Paasivaara, M. and Lassenius, C. (2003).
\newblock {C}ollaboration {P}ractices in {G}lobal {I}nter-{O}rganizational
  {S}oftware {D}evelopment {P}rojects.
\newblock {\em Software Process Improvement and Practice},
  8(4):\text{183--199}.

\bibitem[Parraguez et~al., 2015]{Parraguez15}
Parraguez, P., Eppinger, S.~D., and Maier, A.~M. (2015).
\newblock {I}nformation {F}low {T}hrough {S}tages of {C}omplex {E}ngineering
  {D}esign {P}rojects: {A} {D}ynamic {N}etwork {A}nalysis {A}pproach.
\newblock {\em IEEE Transactions on Engineering Management}, 62(4):604--617.

\bibitem[Perez et~al., 2017]{RodriguezPerez17}
Perez, G.~R., Robles, G., and Barahona, J. M.~G. (2017).
\newblock {H}ow {M}uch {T}ime {D}id {I}t {T}ake to {N}otify a {B}ug? {T}wo
  {C}ase {S}tudies: {E}lastic{S}earch and {N}ova.
\newblock In {\em Proceedings of the IEEE/ACM 8th Workshop on Emerging Trends
  in Software Metrics (WETSoM 2017)}, pages 29--35, Buenos Aires. IEEE.

\bibitem[{Poo-Caama\~{n}o} et~al., 2017]{PooCaamano17}
{Poo-Caama\~{n}o}, G., Knauss, E., Singer, L., and German, D.~M. (2017).
\newblock {H}erding {C}ats in a {FOSS} {E}cosystem: {A} {T}ale of
  {C}ommunication and {C}oordination for {R}elease {M}anagement.
\newblock {\em Journal of Internet Services and Applications}, 8(1):1--24.

\bibitem[Ring, 2015]{Ring15}
Ring, T. (2015).
\newblock {W}hite {H}ats {V}ersus {V}endors: {T}he {F}ight {G}oes {O}n.
\newblock {\em Computer Fraud \& Security}, (10):12--17.

\bibitem[Romo et~al., 2014]{Romo14}
Romo, B.~A., Capiluppi, A., and Hall, T. (2014).
\newblock {F}illing the {G}aps of {D}evelopment {L}ogs and {B}ug {I}ssue
  {D}ata.
\newblock In {\em Proceedings of The International Symposium on Open
  Collaboration (OpenSym 2014)}, pages 1--4, Berlin. ACM.

\bibitem[Ruohonen, 2017]{Ruohonen17TIR}
Ruohonen, J. (2017).
\newblock {C}lassifying {W}eb {E}xploits with {T}opic {M}odeling.
\newblock In {\em Proceedings of the 28th International Workshop on Database
  and Expert Systems Applications (DEXA 2017)}, pages 93--97, Lyon. IEEE.

\bibitem[Ruohonen, 2019]{Ruohonen19ACI}
Ruohonen, J. (2019).
\newblock {A} {L}ook at the {T}ime {D}elays in {CVSS} {V}ulnerability
  {S}coring.
\newblock {\em Applied Computing and Informatics}, 15(2):129--135.

\bibitem[Ruohonen and Allodi, 2018]{Ruohonen18WEIS}
Ruohonen, J. and Allodi, L. (2018).
\newblock {A} {B}ug {B}ounty {P}erspective on the {D}isclosure of {W}eb
  {V}ulnerabilities.
\newblock In {\em Proceedings of the 17th Annual Workshop on the Economics of
  Information Security (WEIS 2018)}, pages 1--14, Innsbruck.
\newblock Available online in June 2019:
  \url{https://weis2018.econinfosec.org/wp-content/uploads/sites/5/2018/05/WEIS_2018_paper_33.pdf}.

\bibitem[Ruohonen et~al., 2016a]{Ruohonen16AICCSA}
Ruohonen, J., Holvitie, J., Hyrynsalmi, S., and Lepp\"anen, V. (2016a).
\newblock {E}xploring the {C}lustering of {S}oftware {V}ulnerability
  {D}isclosure {N}otifications {A}cross {S}oftware {V}endors.
\newblock In {\em Proceedings of the 13th ACS/IEEE International Conference on
  Computer Systems and Applications (AICCSA 2016)}, pages 1--8, Agadir. IEEE.

\bibitem[Ruohonen et~al., 2015]{Ruohonen15COSE}
Ruohonen, J., Hyrynsalmi, S., and Lepp\"anen, V. (2015).
\newblock {T}he {S}igmoidal {G}rowth of {O}perating {S}ystem {S}ecurity
  {V}ulnerabilities: {A}n {E}mpirical {R}evisit.
\newblock {\em Computers \& Security}, 55:1--20.

\bibitem[Ruohonen et~al., 2016b]{Ruohonen16RCIS}
Ruohonen, J., Hyrynsalmi, S., and Lepp\"anen, V. (2016b).
\newblock {T}rading {E}xploits {O}nline: {A} {P}reliminary {C}ase {S}tudy.
\newblock In {\em Proceedings of the IEEE Tenth International Conference on
  Research Challenges in Information Science (RCIS 2016)}, pages 1--12,
  Grenoble. IEEE.

\bibitem[Ruohonen et~al., 2017a]{Ruohonen17COMSIS}
Ruohonen, J., Hyrynsalmi, S., and Lepp\"anen, V. (2017a).
\newblock {M}odeling the {D}elivery of {S}ecurity {A}dvisories and {CVEs}.
\newblock {\em Computer Science and Information Systems}, 14(2):537--555.

\bibitem[Ruohonen and Lepp\"anen, 2017]{Ruohonen17CIT}
Ruohonen, J. and Lepp\"anen, V. (2017).
\newblock {I}nvestigating the {A}gility {B}ias in {DNS} {G}raph {M}ining.
\newblock In {\em Proceedings of the 17th IEEE International Conference on
  Computer and Information Technology (IEEE CIT 2017)}, pages 253--260,
  Helsinki. IEEE.

\bibitem[Ruohonen et~al., 2017b]{Ruohonen17IWSMMensura}
Ruohonen, J., Rauti, S., Hyrynsalmi, S., and Lepp\"anen, V. (2017b).
\newblock {M}ining {S}ocial {N}etworks of {O}pen {S}ource {CVE} {C}oordination.
\newblock In {\em Proceedings of the 27th International Workshop on Software
  Measurement and 12th International Conference on Software Process and Product
  Measurement (IWSM Mensura 2017)}, pages 176--188, Gothenburg. ACM.

\bibitem[Ruohonen et~al., 2016c]{Ruohonen16MESSA}
Ruohonen, J., \v{S}\'{c}epanovi\'{c}, S., Hyrynsalmi, S., Mishkovski, I., Aura,
  T., and Lepp\"anen, V. (2016c).
\newblock {C}orrelating {F}ile-{B}ased {M}alware {G}raphs {A}gainst the
  {E}mpirical {G}round {T}ruth of {DNS} {G}raphs.
\newblock In {\em Proceedings of the 10th European Conference on Software
  Architecture Workshops (ECSAW 2016)}, pages 30:1 -- 30:6, Copenhagen. ACM.

\bibitem[Sabottke et~al., 2015]{Sabottke15}
Sabottke, C., Suciu, O., and Dumitra\c{s}, T. (2015).
\newblock {V}ulnerability {D}isclosure in the {A}ge of {S}ocial {M}edia:
  {E}xploiting {T}witter for {P}redicting {R}eal-{W}orld {E}xploits.
\newblock In {\em Proceedings of the 24th USENIX Security Symposium}, pages
  1041--1056, Washington. USENIX.

\bibitem[Schmid et~al., 2014]{Schmid14}
Schmid, M.~R., Iqbal, F., and Fung, B. C.~M. (2014).
\newblock {E}-{M}ail {A}uthorship {A}ttribution {U}sing {C}ustomized
  {A}ssociative {C}lassification.
\newblock {\em Digital Investigation}, 14(S1):\text{S116--S126}.

\bibitem[Schoch et~al., 2017]{Schoch17}
Schoch, D., Valente, T.~W., and Brandes, U. (2017).
\newblock {C}orrelations {A}mong {C}entrality {I}ndices and a {C}lass of
  {U}niquely {R}anked {G}raphs.
\newblock {\em Social Networks}, 50:46--54.

\bibitem[Seifried, 2017]{Seifried17}
Seifried, K. (2017).
\newblock {CVE}-{HOWTO}.
\newblock Available online in December 2017:
  \url{https://github.com/RedHatProductSecurity/CVE-HOWTO}.

\bibitem[Sierra et~al., 2018]{Sierra18}
Sierra, J.~M., Vizca\'ino, A., Genero, M., and Piattini, M. (2018).
\newblock {A} {S}ystematic {M}apping {S}tudy about {S}ocio-{T}echnical
  {C}ongruence.
\newblock {\em Information and Software Technology}, 94:111--129.

\bibitem[\'{S}liwerski et~al., 2005]{Sliwerski05}
\'{S}liwerski, J., Zimmermann, T., and Zeller, A. (2005).
\newblock {W}hen {D}o {C}hanges {I}nduce {F}ixes? ({O}n {F}ridays.).
\newblock In {\em Proceedings of the International Workshop on Mining Software
  Repositories (MSR 2005)}, pages 1--5, Saint Louis. ACM.

\bibitem[Stevanovic et~al., 2016]{Stevanovic16}
Stevanovic, M., Pedersen, J.~M., {D'Alconzo}, A., and Ruehrup, S. (2016).
\newblock {A} {M}ethod for {I}dentifying {C}ompromised {C}lients {B}ased on
  {DNS} {T}raffic {A}nalysis.
\newblock {\em International Journal of Information Security}, 16(2):115--132.

\bibitem[Syed et~al., 2018]{Syed18}
Syed, R., Rahafrooz, M., and Keisler, J.~M. (2018).
\newblock {W}hat {I}t {T}akes to {G}et {R}etweeted: {A}n {A}nalysis of
  {S}oftware {V}ulnerability {M}essages.
\newblock {\em Computers in Human Behavior}, 80:207--215.

\bibitem[Tang et~al., 2014]{Tang14}
Tang, G., Pei, J., and Luk, W. (2014).
\newblock {E}mail {M}ining: {T}asks, {C}ommon {T}echniques, and {T}ools.
\newblock {\em Knowledge and Information Systems}, 41(1):\text{1--31}.

\bibitem[Tang et~al., 2019]{Tang19}
Tang, M., Alazab, M., and Luo, X. (2019).
\newblock {B}ig {D}ata for {C}ybersecurity: {V}ulnerability {D}isclosure
  {T}rends and {D}ependencies.
\newblock {\em IEEE Transactions on Big Data}, 5(3):317--329.

\bibitem[Temizkan et~al., 2012]{Temizkan12}
Temizkan, O., Kumar, R.~L., Park, S., and Subramaniam, C. (2012).
\newblock {P}atch {R}elease {B}ehaviors of {S}oftware {V}endors in {R}esponse
  to {V}ulnerabilities: {A}n {E}mpirical {A}nalysis.
\newblock {\em Journal of Management of Information Systems}, 28(4):305--337.

\bibitem[Tian et~al., 2016]{TianHassan16}
Tian, Y., Ali, N., Lo, D., and Hassan, A.~E. (2016).
\newblock {O}n the {U}nreliability of {B}ug {S}everity {D}ata.
\newblock {\em Empirical Software Engineering}, 21(6):2298--2323.

\bibitem[Toral et~al., 2009]{Toral09}
Toral, S.~L., {Mart\'inez-Torres}, M.~R., and Barrero, F. (2009).
\newblock {V}irtual {C}ommunities as a {R}esource for the {D}evelopment of
  {OSS} {P}rojects: {T}he {C}ase of {L}inux {P}orts to {E}mbedded {P}rocessors.
\newblock {\em Behavior \& Information Technology}, 28(5):405--419.

\bibitem[Toral et~al., 2010]{Toral10}
Toral, S.~L., {Mart\'inez-Torres}, M.~R., and Barrero, F. (2010).
\newblock {A}nalysis of {V}irtual {C}ommunities {S}upporting {OSS} {P}rojects
  {U}sing {S}ocial {N}etwork {A}nalysis.
\newblock {\em Information and Software Technology}, 52(3):296--303.

\bibitem[Tsipenyuk et~al., 2005]{Tsipenyuk05}
Tsipenyuk, K., Chess, B., and McGraw, G. (2005).
\newblock {S}even {P}ernicious {K}ingdoms: {A} {T}axonomy of {S}oftware
  {S}ecurity {E}rrors.
\newblock {\em IEEE Security \& Privacy}, 3(6):81--84.

\bibitem[Wang, 2014]{Wang14}
Wang, Q. (2014).
\newblock {L}ink {P}rediction and {T}hreads in {E}mail {N}etworks.
\newblock In {\em Proceedings of the International Conference on Data Science
  and Advanced Analytics (DSAA 2014)}, pages \text{470--476}, Shanghai. IEEE.

\bibitem[Wijayasekara et~al., 2012]{Wijayasekara12}
Wijayasekara, D., Manic, M., Wright, J.~L., and McQueen, M. (2012).
\newblock {M}ining {B}ug {D}atabases for {U}nidentified {S}oftware
  {V}ulnerabilities.
\newblock In {\em Proceedings of the 5th International Conference on Human
  System Interactions (HSI 2012)}, pages 89--96, Perth. IEEE.

\bibitem[Wohlin et~al., 2012]{Wohlin12}
Wohlin, C., Runeson, P., H\"ost, M., Ohlsson, M.~C., Regnell, B., and
  Wessl\'en, A. (2012).
\newblock {\em {E}xperimentation in {S}oftware {E}ngineering}.
\newblock Springer, Heidelberg, revised edition.

\bibitem[Wolf et~al., 2009a]{Wolf09b}
Wolf, T., Schroter, A., Damian, D., and Nguyen, T. (2009a).
\newblock {P}redicting {B}uild {F}ailures {U}sing {S}ocial {N}etwork {A}nalysis
  on {D}eveloper {C}ommunication.
\newblock In {\em Proceedings of the IEEE 31st International Conference on
  Software Engineering (ICSE 2009)}, pages 1--11, Vancouver. IEEE.

\bibitem[Wolf et~al., 2009b]{Wolf09a}
Wolf, T., Schr\"oter, A., Damian, D., Panjer, L.~D., and Nguyen, T.~H. (2009b).
\newblock {M}ining {T}ask-{B}ased {S}ocial {N}etworks to {E}xplore
  {C}ollaboration in {S}oftware {T}eams.
\newblock {\em IEEE Software}, 26(1):58--66.

\bibitem[Wu et~al., 2010]{WuGandhi10}
Wu, Y., Gandhi, R.~A., and Siy, H. (2010).
\newblock {U}sing {S}emantic {T}emplates to {S}tudy {V}ulnerabilities
  {R}ecorded in {L}arge {S}oftware {R}epositories.
\newblock In {\em Proceedings of the 2010 ICSE Workshop on Software Engineering
  for Secure Systems (SESS 2010)}, pages 22--28, Cape Town. ACM.

\bibitem[Wu and Oard, 2005]{Wu05}
Wu, Y. and Oard, D.~W. (2005).
\newblock {I}ndexing {E}mails and {E}mail {T}hreads for {R}etrieval.
\newblock In {\em Proceedings of the 28th Annual International ACM SIGIR
  Conference on Research and Development in Information Retrieval (SIGIR
  2005)}, pages \text{665--666}, Salvador. ACM.

\bibitem[Younis et~al., 2016]{YonisMalaiya16}
Younis, A., Malaiya, Y.~K., and Ray, I. (2016).
\newblock {E}valuating {CVSS} {B}ase {S}core {U}sing {V}ulnerability {R}ewards
  {P}rograms.
\newblock In Hoepman, J.-H. and Katzenbeisser, S., editors, {\em Proceedings of
  the 31st IFIP TC 11 International Conference on ICT Systems Security and
  Privacy Protection (IFIP SEC 2016)}, pages 62--75, Ghent. Springer.

\bibitem[Zanetti et~al., 2013a]{Zanetti13a}
Zanetti, M.~S., Scholtes, I., Tessone, C.~J., and Schweitzer, F. (2013a).
\newblock {C}ategorizing {B}ugs with {S}ocial {N}etworks: {A} {C}ase {S}tudy on
  {F}our {O}pen {S}ource {S}oftware {C}ommunities.
\newblock In {\em Proceedings of the 35th International Conference on Software
  Engineering (ICSE 2013)}, pages 1032--1041, San Francisco. IEEE.

\bibitem[Zanetti et~al., 2013b]{Zanetti13b}
Zanetti, M.~S., Scholtes, I., Tessone, C.~J., and Schweitzer, F. (2013b).
\newblock {T}he {R}ise and {F}all of a {C}entral {C}ontributor: {D}ynamics of
  {S}ocial {O}rganization and {P}erformance in the {GENTOO} {C}ommunity".
\newblock In {\em Proceedings of the 6th International Workshop on Cooperative
  and Human Aspects of Software Engineering (CHASE 2013)}, pages 49--56, San
  Francisco. IEEE.

\bibitem[Zangerle et~al., 2013]{Zangerle13}
Zangerle, E., Gassler, W., and Specht, G. (2013).
\newblock {O}n the {I}mpact of {T}ext {S}imilarity {F}unctions on {H}ashtag
  {R}ecommendations in {M}icroblogging {E}nvironments.
\newblock {\em Social Network Analysis and Mining}, 3(4):\text{889--898}.

\bibitem[Zeileis, 2004]{Zeileis04}
Zeileis, A. (2004).
\newblock {E}conometric {C}omputing with {HC} and {HAC} {C}ovariance {M}atrix
  {E}stimators.
\newblock {\em Journal of Statistical Software}, 11(10):\text{1--17}.

\bibitem[Zhang et~al., 2012]{ZhangHassan12}
Zhang, F., Khomh, F., Zou, Y., and Hassan, A.~E. (2012).
\newblock {A}n {E}mpirical {S}tudy on {F}actors {I}mpacting {B}ug {F}ixing
  {T}ime.
\newblock In {\em Proceedings of the 19th Working Conference on Reverse
  Engineering (WCRE 2012)}, pages 225--234, Kingston. IEEE.

\bibitem[Zhou et~al., 2015]{ZhouGupta15}
Zhou, B., Neamtiu, I., and Gupta, R. (2015).
\newblock {E}xperience {R}eport: {H}ow {D}o {B}ug {C}haracteristics {D}iffer
  {A}cross {S}everity {C}lasses: {A} {M}ulti-{P}latform {S}tudy.
\newblock In {\em Proceedings of the 26th International Symposium on Software
  Reliability Engineering (ISSRE 2015)}, pages 507--517, Gaithersbury. IEEE.

\end{thebibliography}

\end{document}